\documentclass[]{aa}  

\usepackage{graphicx}
\usepackage{txfonts}
\usepackage{lscape}
\usepackage{caption} %For continuing figure 7 over multiple pages
\usepackage{bm} %For bolding terms in equations
\usepackage{float}
\newcommand{\degsq}{deg$^{2}$\,}
\usepackage[]{hyperref}
\begin{document}

   \title{The XXL Survey: \\ XXXIX. Polarised radio sources in the XXL-South field}

   \author{R. A. J. Eyles\inst{1}\thanks{Corresponding author}
          \and
          M. Birkinshaw \inst{2}
        \and
        V. Smol{\v c}i{\'c}\inst{3}
        \and
        C. Horellou\inst{4}
        \and 
        M. Huynh\inst{5,6}
        \and
        A. Butler\inst{5}
        \and J. Delhaize\inst{7}
        \and C. Vignali\inst{8,9}
        \and M. Pierre\inst{10}
          }

   \institute{Department of Physics and Astronomy, University of Leicester, University Road, Leicester, LE1 7RH, UK
              \email{raje1@leicester.ac.uk}
         \and
             HH Wills Physics Laboratory, University of Bristol, Tyndall Avenue, Bristol, BS8 1TL, UK
             \and
             Physics Department, University of Zagreb, Bijeni{\v c}ka cesta 32, 10002 Zagreb, Croatia
            \and
            Department of Space, Earth and Environment, Chalmers University of Technology, Onsala Space Observatory, SE-439 92 Onsala, Sweden
            \and
            International Centre for Radio Astronomy Research (ICRAR), University of Western Australia, 35 Stirling Hwy, Crawley WA 6009, Australia
            \and
            CSIRO Astronomy and Space Science, 26 Dick Perry Ave, Kensington WA 6151, Australia
            \and
            Department of Astronomy, University of Cape Town, Private Bag X3, Rondebosch 7701, South Africa
            \and
            Dipartimento di Fisica e Astronomia, Alma Mater Studiorum, Università degli Studi di Bologna, Via Gobetti 93/2, 40129 Bologna, Italy
            \and
            INAF – Osservatorio di Astrofisica e Scienza dello Spazio di Bologna, Via Gobetti 93/3, 40129 Bologna, Italy
            \and
            AIM, CEA, CNRS, Universit\'{e} Paris-Saclay, Universit\'{e} Paris Diderot, Sorbonne Paris Cit\'{e}, F-91191 Gif-sur-Yvette, France
             }

   \date{Received <date>; accepted <date>}
   
   \titlerunning{XXL XXXIX: Polarised radio sources in the XXL-South field% and leakage in ATCA mosaics
   }

% \abstract{}{}{}{}{} 
% 5 {} token are mandatory
 
  \abstract
   {}
   {We investigate the properties of the polarised radio population in the central 6.5 deg$^{2}$ of the XXL-South field observed at 2.1 GHz using the Australia Telescope Compact Array (ATCA) in 81 pointings with a synthesised beam of FWHM 5.2\arcsec. We also investigate the ATCA's susceptibility to polarisation leakage.}
   {We performed a survey of a 5.6 deg$^{2}$ subregion and calculated the number density of polarised sources. We derived the total and polarised spectral indices, in addition to comparing our source positions with those of X-ray-detected clusters. We measured the polarisation of sources in multiple pointings to examine leakage in the ATCA.}
   {We find 39 polarised sources , involving 50 polarised source components, above a polarised flux density limit of 0.2 mJy at 1.332 GHz. The number density of polarised source components is comparable with recent surveys, although there is an indication of an excess at $\sim1$ mJy. We find that those sources coincident with X-ray clusters are consistent in their properties with regard to the general population. In terms of the ATCA leakage response, we find that ATCA mosaics with beam separation of $\lesssim 2/3$ of the primary beam FWHM have off-axis linear polarisation leakage $\lesssim 1.4$\% at 1.332 GHz.}
   {}

               
\defcitealias{Pierre16}{XXL~Paper~I}
\defcitealias{Pacaud16}{XXL~Paper~II}
\defcitealias{Giles16}{XXL~Paper~III}
\defcitealias{Fotopoulou16}{XXL~Paper~VI}
\defcitealias{Smolcic16}{XXL~Paper~XI}
\defcitealias{Lidman16}{XXL~Paper~XIV}
\defcitealias{Butler18a}{XXL~Paper~XVIII}
\defcitealias{Adami18}{XXL~Paper~XX}
\defcitealias{Chiappetti18}{XXL~Paper~XXVII}
\defcitealias{Butler18b}{XXL~Paper~XXXI}

\maketitle

%
%________________________________________________________________

\section{Introduction}
\label{sec:Intro}

Characterising the polarised source population is particularly important as the completion of the Square Kilometre Array (SKA) approaches. The cumulative number density of this population determines the number of polarised sources detectable by the SKA. The population of distant polarised sources is key in examining foreground cosmic magnetic fields using Faraday rotation \citep{Faraday46}, a core science objective of the SKA \citep[e.g.][]{Beck04,Johnston15,Bonafede15}. 

In addition, polarised source counts can be used to complement total intensity counts to determine source properties \citep[e.g.][]{Law11,Guidetti12} and characterise the overall source population. This is crucial in, for example, examining the relationship between AGN activity and galaxy evolution \citep[e.g.][]{McAlpine15} or when morphologically classifying galaxies \citep[e.g.][]{Makhathini15}, both of which are also core science objectives of the SKA.

Here we investigate source polarisation in radio sources in the XXL survey, which covers two 25 deg$^{2}$ fields, one equatorial field (XXL-North) and one in the southern hemisphere (XXL-South) \citep[hereafter \citetalias{Pierre16}]{Pierre16}. As part of follow-up work for the XXL survey, a 6.5 deg$^{2}$ region within the XXL South field was observed in 81 pointings at 2.1 GHz using the Australia Telescope Compact Array (ATCA) \citep[hereafter \citetalias{Smolcic16}]{Smolcic16}. This was carried out as a pilot for observations of the full XXL South field \citep[hereafter \citetalias{Butler18a}]{Butler18a}. While total intensity data reduction and imaging of the field have been published in \citetalias{Smolcic16}, the data offer an opportunity to examine certain properties of the polarised source population, such as their number density and spectral indices.

The particular field examined here overlaps with one of the DASI fields examined by \citet{Bernardi06} (their Field 1). \citeauthor{Bernardi06} present 1.4 GHz ATCA measurements of the total intensity, polarised intensity and polarisation angle for 18 sources at an angular resolution of 3.4$\arcmin$ and a noise of 18 mJy~beam$^{-1}$. A number of these sources are common to the data presented here and allow a comparison. 
The dataset also allows for an examination of the ATCA's polarisation leakage. Leakage comes about as the result of an unintended response by the linear polarisation feed systems to the wrong polarisation, that is, the Stokes $Q$ feed system responding to Stokes $U$ and vice versa. Alternatively, it can be considered as Stokes $I$ leaking into $Q$ and $U$. This response varies depending on the relative positioning of the pointing centre and the source, and it changes with frequency. It is possible to monitor leakage by examining how the $Q$, $U$ and, to a lesser extent, Stokes $V$, vary for a source with known polarisation being observed from many positions. Because the leakage terms are expected to vary while the inherent polarisation of the source remains constant, it should be possible to extract a telescope-specific correction function which can be applied to future observations. Previous work has been undertaken to identify the correction functions of other arrays, such as the Murchison Widefield Array \citep[MWA;][]{Sutinjo15}, and the ATCA's frequency response to on-axis leakage is well-known. However, a comprehensive solution in the context of mosaic observations has not yet been identified. By comparing the data from individual pointings within our dataset, it is possible to extract information on the ATCA's leakage response and identify some general characteristics of the correction function.

Sections 2 and 3 summarise our observations, data reduction, and imaging. In section 4, the properties of the polarised source population are examined and contrasted with the findings of \citet{Rudnick14a,Rudnick14b} while section 5 deals with our investigation into the leakage present in ATCA mosaics. Our conclusions are summarised in section 6.

\section{Observations}

Observations were performed with the ATCA, a configurable array of six 22 m antennas operated by the Australia Telescope National Facility (ATNF). The ATCA is an earth-rotation aperture synthesis radio interferometer \citep{Stevens18}.

Two sets of observations were performed. The first was conducted for a period of 37 hours between 3-6 September 2012. For this first session, the antennas were arranged in the 6A configuration. A second set of observations was performed using the 1.5C configuration over 15 hours betwen 25-26 November 2012 \citepalias{Smolcic16}.

81 mosaic pointings cover the 6.5 \degsq field. These pointings were placed so that the separations in both RA and Dec were 2/3 of the primary beam FWHM, which is 14.7 arcmin at the central observing frequency of 2.1 GHz.

The primary calibrator was PKS 1934-638, a long-established calibrator with well-known properties \citep{Reynolds94}. It was observed on-source during each observing run for ten minutes. The secondary calibrator was PKS 2333-528, which was observed for two minutes on-source between 32-minute observations of different sets of pointings.

\section{Data reduction and imaging}

\begin{figure*}
\centering
\includegraphics[width=180mm]{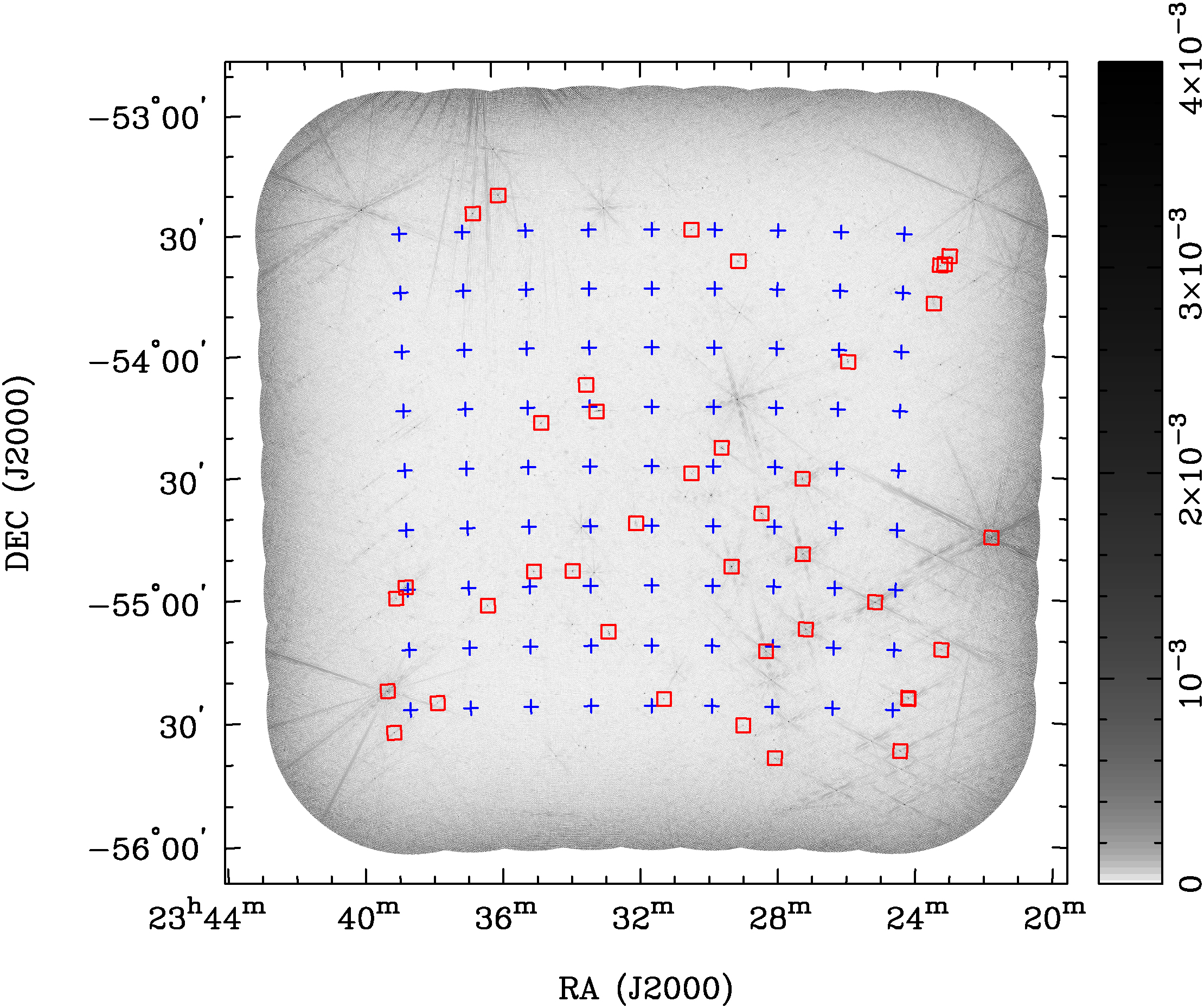}
\caption{Total intensity mosaic at 1332 MHz. Sources in the XXL$_{\text{39}}$ dataset are indicated with red boxes and centres of each pointing are indicated with blue crosses. Grey scale is in Jy~beam$^{-1}$ and the FWHM of the synthesised beam is 5.2\arcsec.}
\label{fig:IMosaic}
\end{figure*}

\begin{figure*}
\centering
\includegraphics[width=180mm]{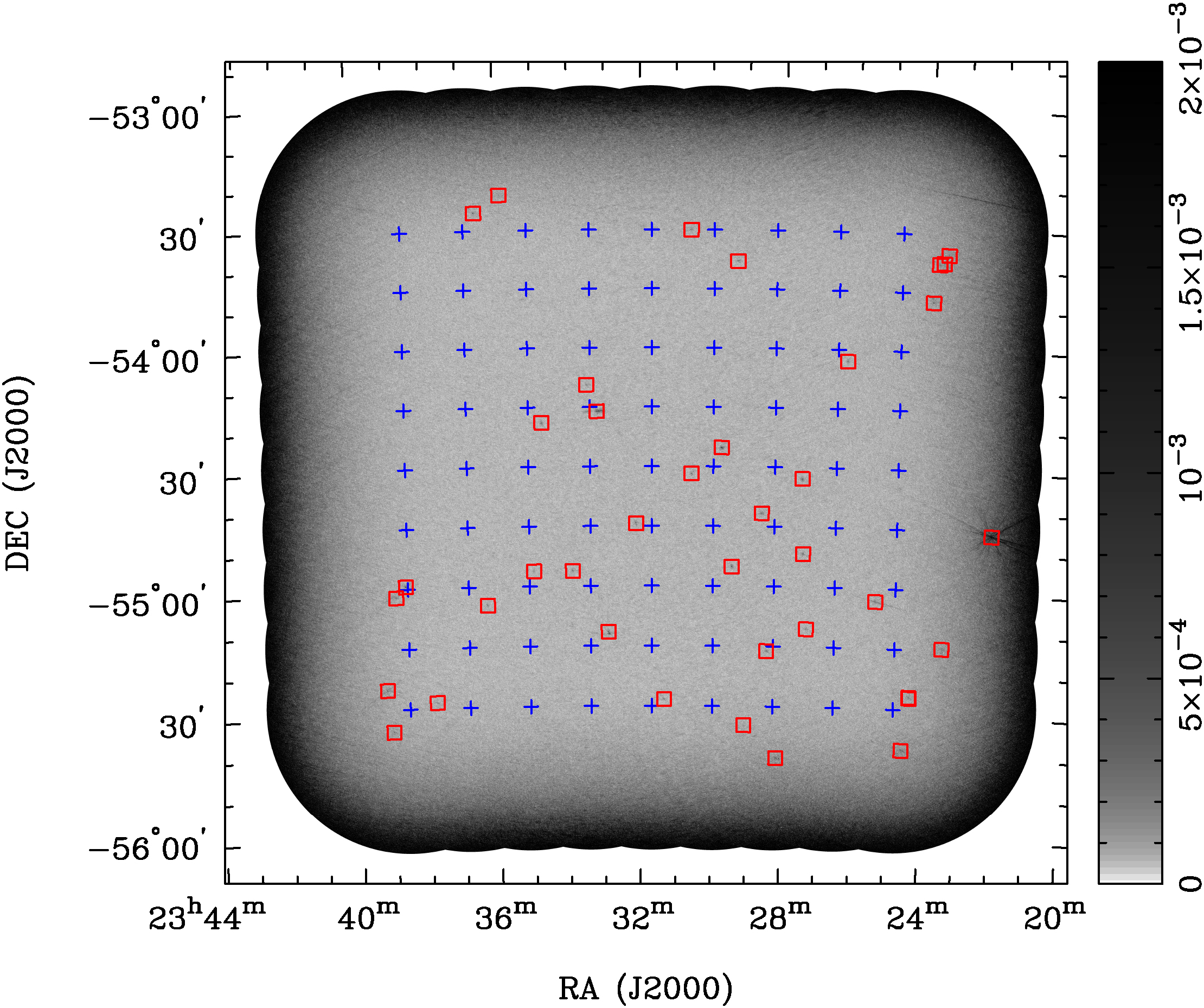}
\caption{As in Figure \ref{fig:IMosaic}, but showing the polarised intensity mosaic at 1332 MHz instead.}
\label{fig:PMosaic}
\end{figure*}

Data reduction and imaging were performed using the \textsc{Miriad} (Multichannel Image Reconstruction, Image Analysis and Display) software \citep{Sault95}. 

16 frequency bands of width 128 MHz were used in the calibration step, performed using the \textsc{Miriad} task \texttt{gpcal} (gain/phase/polarisation calibration).

Automatic and manual flagging were performed using the \texttt{pgflag} and \texttt{blflag} tasks in \textsc{Miriad}. Following multiple iterations of these, an average of $20.7\pm 2.2\%$ of the raw data were flagged for each pointing, with generally more data being flagged in the pointings at lower elevations. The time of the observation appeared to have little effect on the amount of flagging.

A common problem with wideband imaging is the varying frequency response by the receiver, in this case the Compact Array Broadband Backend or CABB \citep{Wilson11}. In order to minimise this effect on the results, the data were split into four 512 MHz wide wavebands centred at 1332, 1844, 2356, and 2868 MHz. The data for each individual pointing and for each Stokes parameter were reduced and imaged separately. The visibilities at each of these bands were also given a robust weighting with a Briggs parameter \citep{Briggs95} of 0.5 to ensure the synthesised beam size was broadly consistent across the full dataset.

The image size was set at 16384 by 16384 pixels as the \textsc{Miriad} task \texttt{mfclean}, which implements a multi-frequency version of the \texttt{clean} algorithm, only works on the central part of the image. The large size, therefore, ensured that the sidelobes would be sufficiently cleaned. Self-calibration was found to be ineffective due to the low signal to noise ratio, which was expected as the XXL Survey fields had been selected in the aim of excluding extremely bright sources \citepalias{Pierre16}.

The images were restored using a  diameter of 5.2\arcsec\   FWHM of the circular Gaussian beam. This is more conservative than the 4.7\arcsec\, by 4.2\arcsec\, beam used by \citetalias{Smolcic16} but it produces comparable results. The final images were combined using the \texttt{linmos} task. A mosaic for each Stokes parameter in each band was produced, as well as a total intensity mosaic averaged across the bands and a polarised intensity mosaic, found from $P=\sqrt{Q^2+U^2}$, for each band. The total and polarised intensity mosaics for the 1332 MHz band are shown in Figures~\ref{fig:IMosaic} and~\ref{fig:PMosaic}, respectively.

In terms of quality, the total intensity mosaics have a mean rms of 53 $\mu$Jy beam$^{-1}$ and the polarised intensity mosaics have a mean rms of 35 $\mu$Jy beam$^{-1}$. The mosaics are sampled at 1.64 arcsec per pixel. The remaining artifacts are caused by unremoved sidelobes. These are difficult to remove due to imperfect antenna calibration and clean modelling, errors which have been minimised in subsequent observations; for instance, by introducing a more rigorous cleaning process, which includes peeling \citepalias{Smolcic16,Butler18a}.

\section{Polarised source population}

\subsection{Identification of polarised sources}
\label{sec:SourceID}

\begin{figure}
\includegraphics[width=\columnwidth]{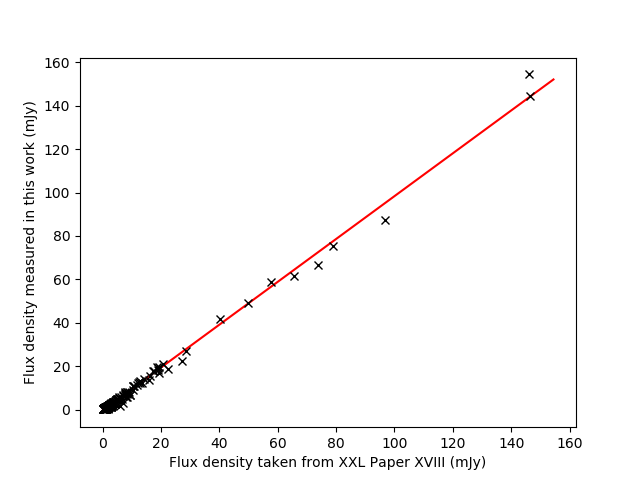}
\caption{Comparison of estimated flux densities for the sources found with our initial source count in the catalogue detailed in \citetalias{Butler18a}. The red line is a linear fit of the data and is consistent with a slope of unity.}
\label{fig:FluxComp}
\end{figure}

A source count was performed on the total intensity mosaic at 2.1 GHz using the \textsc{Miriad} \texttt{sfind} task, which uses the false discovery rate (FDR) method to identify sources \citep{Hopkins02} and then adopts a least-squares routine to fit a 2D elliptical gaussian to the source and measure its peak and integrated intensities \citep{Sault04,Condon97}. A $p$-value threshold of $2.7\times 10^{-5}$ or $4 \sigma$ was selected and a total of 1316 sources were detected, slightly lower than the 1386 found by \citetalias{Smolcic16} in the same field. This is most likely due to more conservative conditions for source detections in our count. We compared our source count to the updated source catalogue detailed in \citetalias{Butler18a}\footnote{\url{http://vizier.u-strasbg.fr/viz-bin/VizieR-2?-source=+IX\%2F52\%2Fatcacomp}} by cross-matching between them using \textsc{Topcat} \citep{Taylor05}. We plotted our flux densities against the catalogue as in Figure~\ref{fig:FluxComp}, dividing their values by 1.035 to account for the bandwidth smearing correction, and finding a slope of 0.99 $\pm$ 0.01 and an intercept of -0.39 $\pm$ 0.01 mJy, indicating that we measure slightly lower, but comparable, flux densities.

A source list was also generated from the 1332 MHz polarised intensity mosaic. The area over which this source count was performed (5.634 \degsq) was smaller than the full mosaic in order to reduce spurious detections from noise at the edge of our mosaic. This frequency band was chosen as it most closely matched 1.4 GHz, as used by \citet{Rudnick14a,Rudnick14b} in their deep Very Large Array (VLA) observations of the GOODS-N field. \texttt{Sfind} still produced a number of spurious sources, so a cross check with our previous total intensity source list, in addition to manual checks, was used to confirm genuine sources. We also set a sensitivity cutoff at 0.2 mJy given that  below this level, errors tended to be larger than the measured flux density. We also corrected for Ricean bias using the solution of \citet{Wardle74} $\left(P\sim\sqrt{P_{\text{obs}}^2-\sigma_P^2}\right)$.

The intrinsically polarised nature of the emission in these sources indicates that it traces the synchrotron radiation induced in AGN jets or radio lobes rather than emission from star-forming galaxies, but as the polarisation percentages are relatively low, it is likely that significant beam depolarisation is present. This final dataset is designated hereafter as the XXL$_{\text{39}}$ dataset. Figures~\ref{fig:IMosaic} and~\ref{fig:PMosaic} show the locations of these 39 sources with polarised intensity $\gid 0.2$ mJy within the field. A number of these sources consist of several components, which were detected as separate sources by \texttt{sfind}. Manual inspection, including comparisons with optical and IR imaging, identified them as parts of the same source. In total, we detected 50 components, including sources consisting of only a single component, and for some of our subsequent analysis, we treated each component separately. Several examples from our source population, including all multi-component sources, are displayed in Figure~\ref{fig:SourceExamples}. Some properties of the sources in the XXL$_{\text{39}}$ dataset are summarised in Table~\ref{tab:SourceList}.

We matched XXL$_{\text{39}}$ to the closest previously-known sources found in the NASA/IPAC Extragalactic Database, NED\footnote{\url{https://ned.ipac.caltech.edu/}} \citep{Mauch03}, finding that only three of our sources had redshift data available. We also compared our source list with the photometric catalogue detailed in \citet[hereafter \citetalias{Fotopoulou16}]{Fotopoulou16}\footnote{\url{http://vizier.u-strasbg.fr/viz-bin/VizieR-3?-source=IX/49/xxl1000a}} and the spectroscopic catalogues detailed in \citet[hereafter \citetalias{Lidman16}]{Lidman16}\footnote{\url{http://vizier.u-strasbg.fr/viz-bin/VizieR-3?-source=IX/49/xxlaaoz}} and \citet[hereafter \citetalias{Chiappetti18}]{Chiappetti18}\footnote{\url{http://vizier.u-strasbg.fr/viz-bin/VizieR-3?-source=IX/52/xxlaaoz}}, identifying redshifts for a further ten sources. In instances where discrepancies between sources are identified, such as for PKS 2319-55 and WISE J233913.22-552350.8,  the spectroscopic redshift measured in \citetalias{Chiappetti18} is favoured. Of the 13 sources with redshifts available, nine are at redshifts lower than 1. \citetalias{Fotopoulou16} and \citetalias{Lidman16} also identify five of these sources as QSOs or AGN but they do not classify the remainder. The available redshift data for our source population are summarised in Table \ref{tab:redshifts}.

\begin{figure}
\includegraphics[width=\columnwidth]{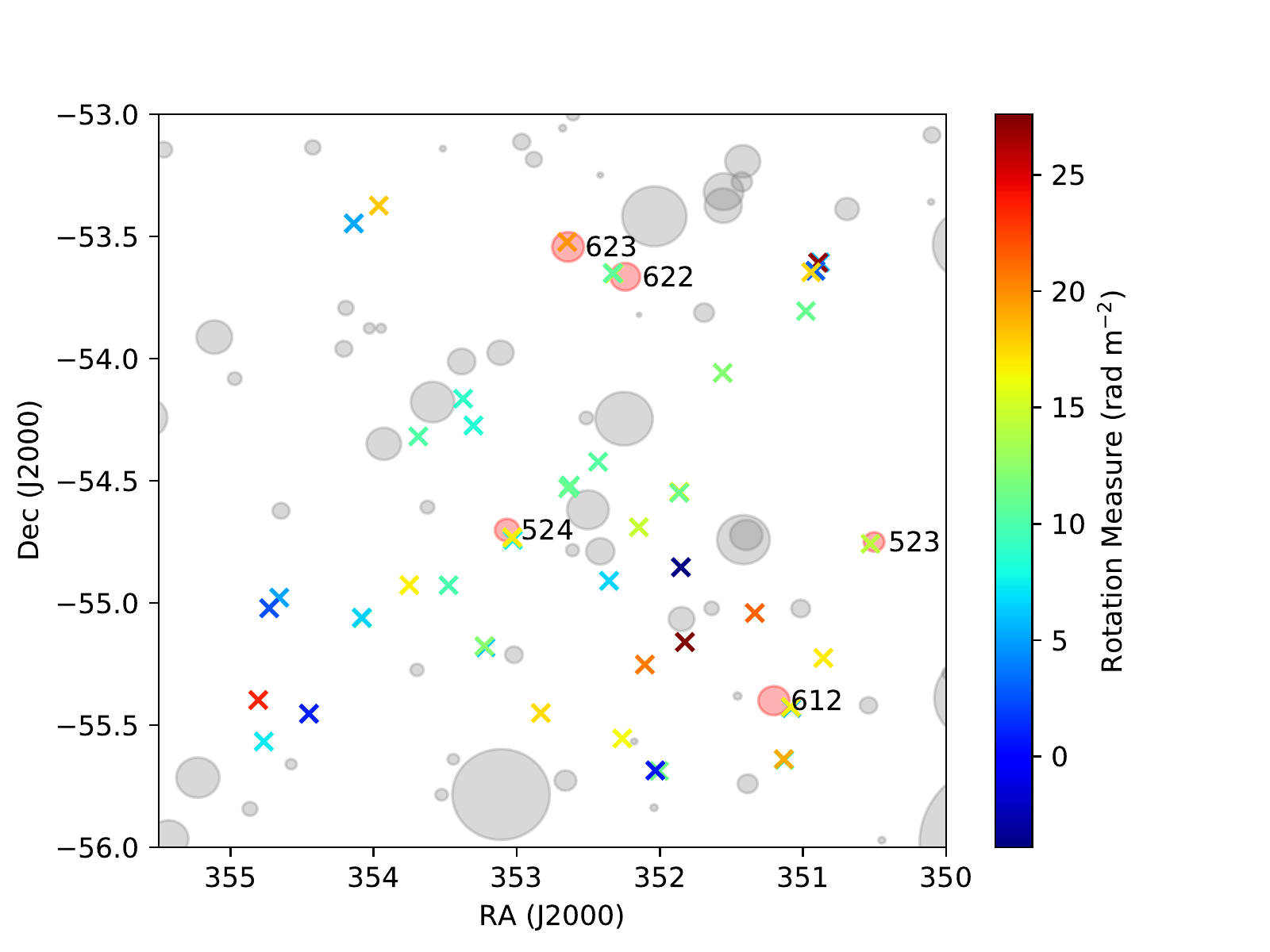}
\caption{XXL$_{39}$ source positions, indicated with crosses, relative to the X-ray identified clusters detailed in \citetalias{Pacaud16}, \citetalias{Giles16}, and \citetalias{Adami18}, indicated with shaded circles with radius equal to their $r_{200}$s. When a source overlaps with a cluster, regardless of the source and the cluster's redshifts, the cluster is shaded in red and its designation from \citetalias{Adami18} is included. The colourbar is used to indicate the rotation measures of the polarised sources.}
\label{fig:ClusterComp}
\end{figure}

We compared our source positions with the positions of the X-ray identified clusters detailed in \citet[hereafter \citetalias{Pacaud16}]{Pacaud16}, \citet[hereafter \citetalias{Giles16}]{Giles16} and \citet[hereafter \citetalias{Adami18}]{Adami18}. As shown in Figure \ref{fig:ClusterComp}, we find six sources that are positioned within, or close to, a cluster's $r_{200}$, the radius at which the local density is 200 times that of the critical density calculated as $r_{200} \approx r_{500} / 0.65$ \citep{Ettori09} when projected onto the sky. Of these sources, two (WISEA J233035.37-533122.5 and 2MASS J23320704-5444040) are at comparable redshifts to their nearest clusters and are plausibly embedded within them. The six sources are detailed in Table \ref{tab:SourceCluster}.

We also note that most of the brightest sources in total intensity have a polarisation percentage ($P/I$) of less than 2\%. This can be expected as only certain classes of AGN, such as blazars and BL Lacs, emit strongly polarised radiation and so they are the dominant contributors in the XXL$_{39}$ dataset. Furthermore, if there are multiple unresolved components within a source, the $Q$ and $U$ measurements will tend to average out, so the integrated polarisation over the whole source will be lower than each individual component. In the present dataset, we expect such beam depolarisation effects to be important for a subset of sources with angular size less than 5 arcsec. Due to the wide-field nature of the observations, it is also likely that the polarised intensities that have been measured include a contribution from off-axis linear polarisation leakage. However, due to the observing strategy and the mosaicing method used, the maximum leakage should be relatively small. Based on previous measurements of the ATCA's leakage response \citep{Sault99,Anderson15}, the maximum leakage for the most off-axis sources should be $\leq 1\%$, and it is, therefore, unlikely that the polarised intensities of the sources in the XXL$_{\text{39}}$ dataset are significantly overestimated. This is reinforced by our own investigation into the leakage response of the ATCA (see Section \ref{sec:Leakage}).

\begin{table*}
\caption{Redshifts for sources in the XXL$_{\text{39}}$ dataset.}         % title of Table
\label{tab:redshifts}      % is used to refer this table in the text
\centering                          % used for centering table
\begin{tabular}{lcccc}        % centered columns (3 columns)
\hline\hline                 % inserts double horizontal lines
Source & $z$ from & $z$ from & $z$ from & $z$ from \\
& NED & \citetalias{Fotopoulou16} & \citetalias{Lidman16} & \citetalias{Chiappetti18}\\    % table heading 
\hline                        % inserts single horizontal line
   PKS 2319-55 & 0.730$^{\textit{1}}$ & 0.878 & 1.064 & 0.878 \\      % inserting body of the table
   MRSS 191-011762 & --- & --- & --- & 0.334 \\
   SSTSL2 J232346.32-533847.1 & --- & 1.961 & 1.995 & ---    \\
   SCSO J232419.6-552548.1 & 0.241$^{\textit{2}}$ & ---  & 0.240 & 0.240   \\
   SSTSL2 J232520.78-550228.8 & --- & 1.110 & 1.526 & 1.524 \\
   SUMSS J232614-540321 & --- & --- & 1.663 & --- \\
   SSTSL2 J232727.30-543250.7 & --- & --- & 0.319 & --- \\
   SSTSL2 J232925.10-545435.7 & --- & --- & --- & 0.468 \\
   WISEA J233035.37-533122.5 & --- & --- & 0.171 & 0.171 \\
   2MASS J23320704-5444040 & --- & --- & 0.273 & 0.273 \\
   SSTSL2 J233619.37-550342.0 & --- & --- & --- & 0.401 \\
   SSTSL2 J233838.02-545841.3 & --- & --- & 0.527 & ---\\
   WISE J233913.22-552350.8 & 1.354$^{\textit{3}}$ & 0.049 & 1.355 & --- \\
\hline %inserts single line
\end{tabular}
\\ \footnotesize{$^{\textit{1}}$Estimated from $R$ band \citep{Burgess06}; $^{\textit{2}}$\citet{Suhada12}; $^{\textit{3}}$\citet{Wisotzki00}.}
\end{table*}

\subsection{Cumulative source count properties}

We compared our polarised source population to that found in a deeper survey using the Very Large Array (VLA). \citet{Rudnick14a,Rudnick14b} report sources with polarised flux density $P\gtrsim 14.5$ $\mu$Jy. The cumulative number density, defined as $N(P>P_0)$/\degsq, where $P$ is the polarised flux density of the source and $P_0$ is the lower bin limit, when combined with results from earlier surveys \citep{Taylor07,Grant10,Subrahmanyan10}, seems to indicate a turnover in the source counts at $P\sim0.6$ mJy. It is possible that leakage effects could have an impact on these results, pushing the flux measurements higher, especially for brighter sources.

\begin{figure}
\includegraphics[width=\columnwidth]{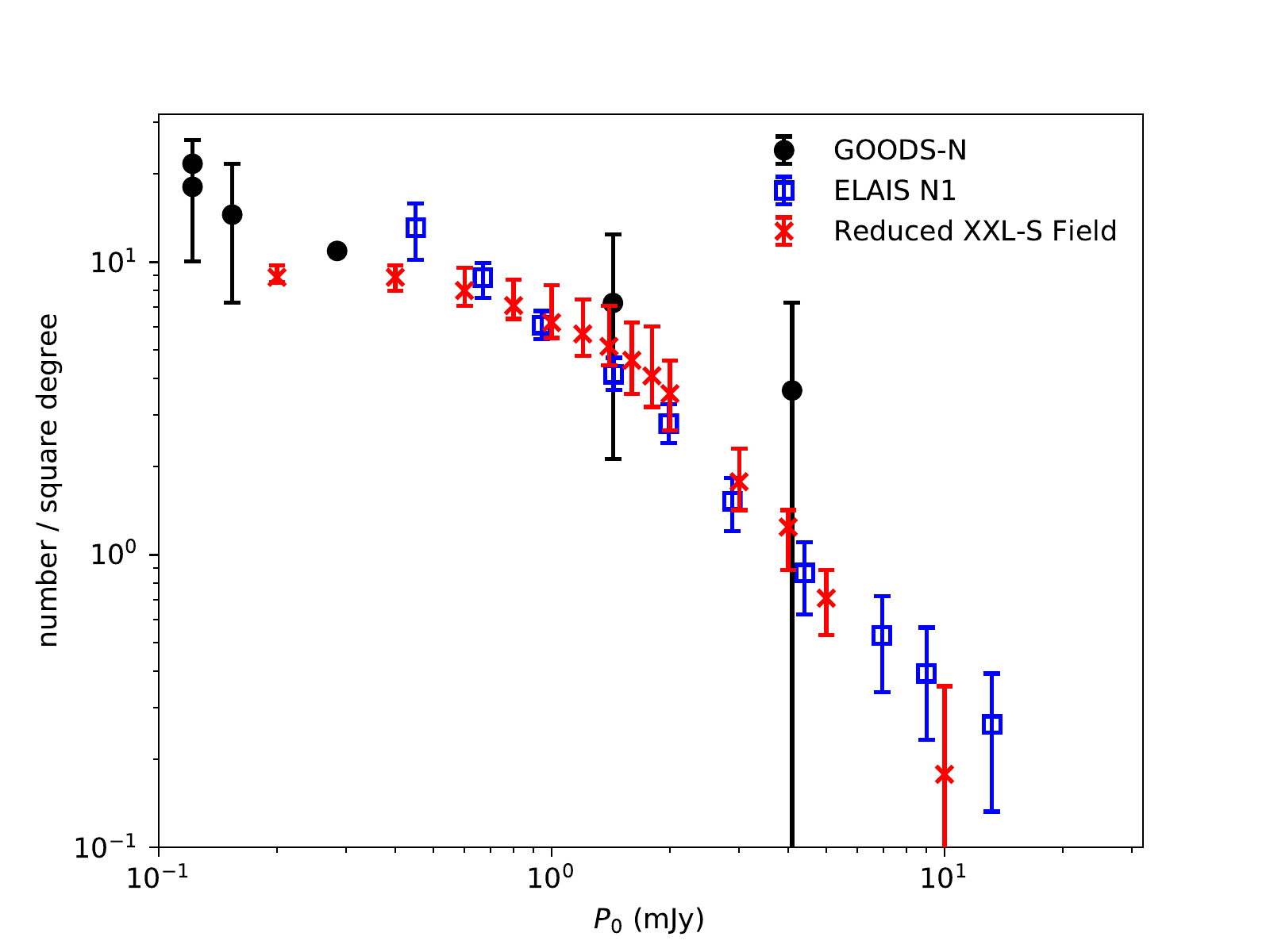}
\caption{Cumulative number density of sources against polarised flux density in the field examined here is shown in red. The results from \citet{Rudnick14a,Rudnick14b} are shown in black. The count from the ELAIS-N1 field \citep{Grant10} is shown in blue.}
\label{fig:VLAComp}
\end{figure}

We calculated the cumulative number density of polarised sources in our field, considering each component of multi-component sources as an individual source. As shown in Figure~\ref{fig:VLAComp}, the minimum flux density for detected polarised sources in our data is $P\sim 0.2$ mJy. Although our survey is substantially shallower than that of \citeauthor{Rudnick14a}, we independently confirm the steepening of the polarised source count at $P\gtrsim 1.5$ mJy,  although there is an indication of a higher number density at $\sim$1.0 to 1.2 mJy compared to the ELAIS N1 results. While the reason for this disparity is unclear, this may imply that \citet{Grant10}'s count deficit is particular to their selected field or that our excess is particular to ours. The more extreme flattening of our count at P$_0$ < 0.4 mJy is due to incompleteness near the flux density limit.

\subsection{Spectral indices and depolarisation}
\label{sec:SpecIndex}

\begin{figure}
\includegraphics[width=\columnwidth]{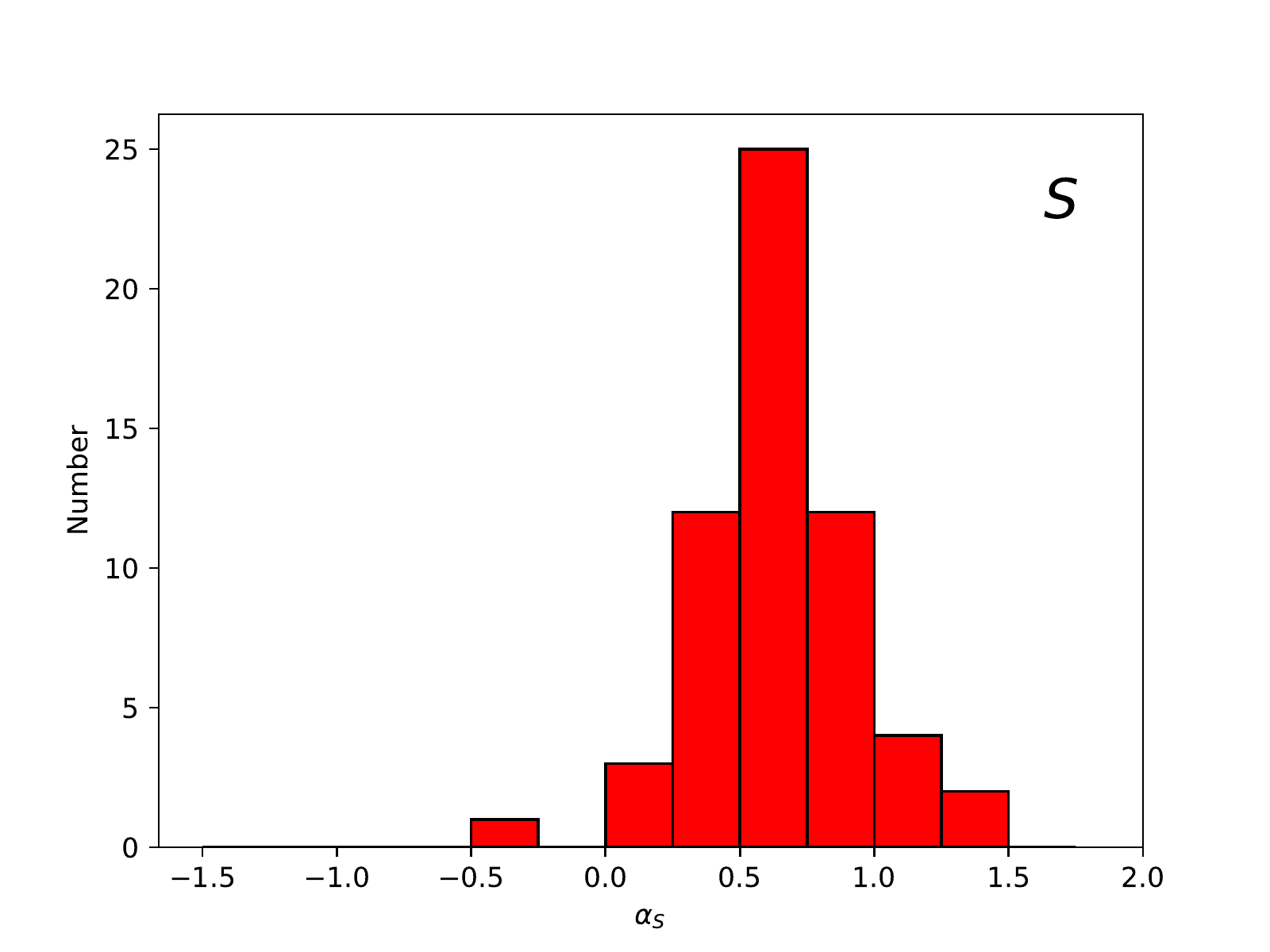}
\includegraphics[width=\columnwidth]{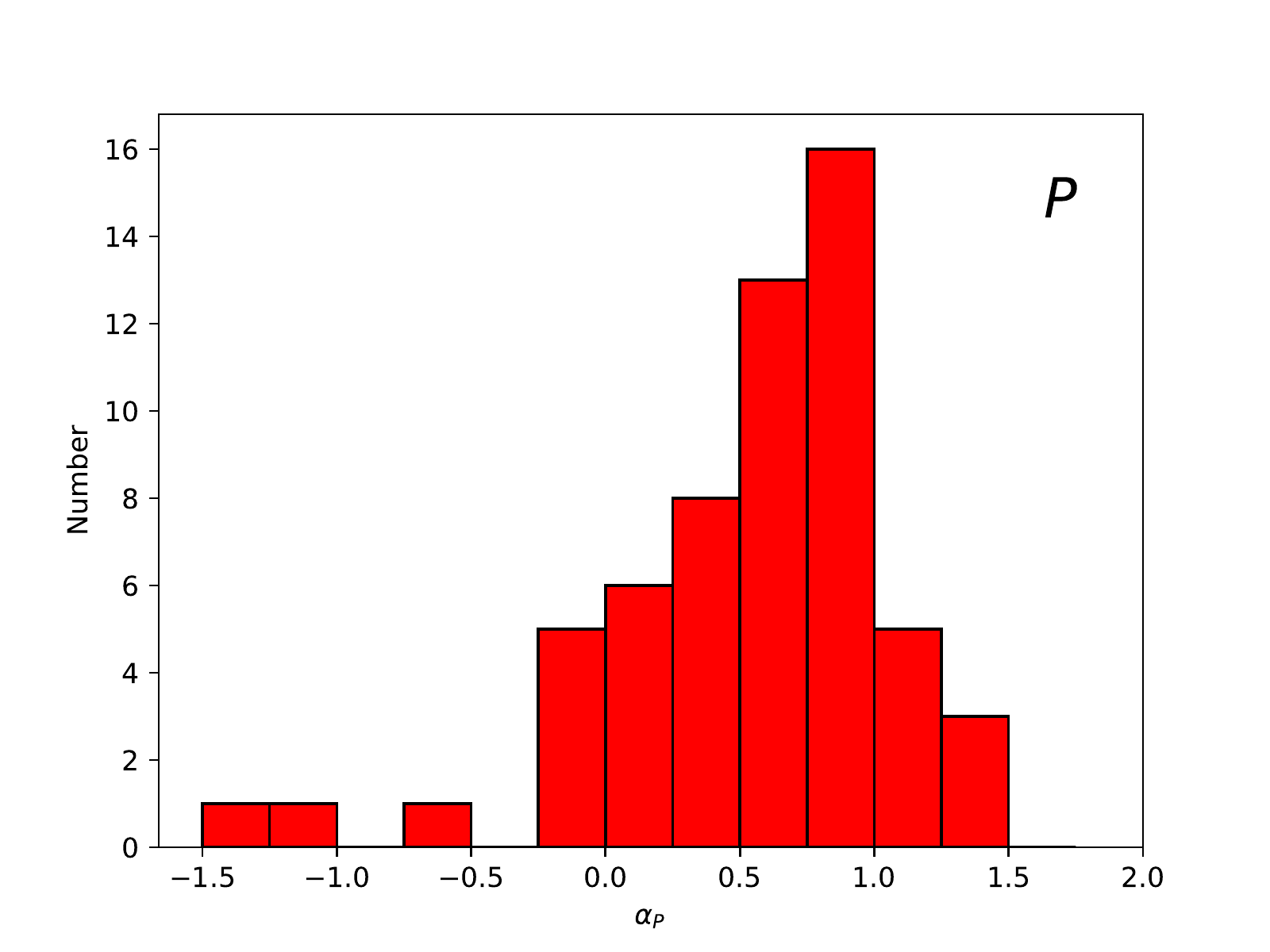}
\includegraphics[width=\columnwidth]{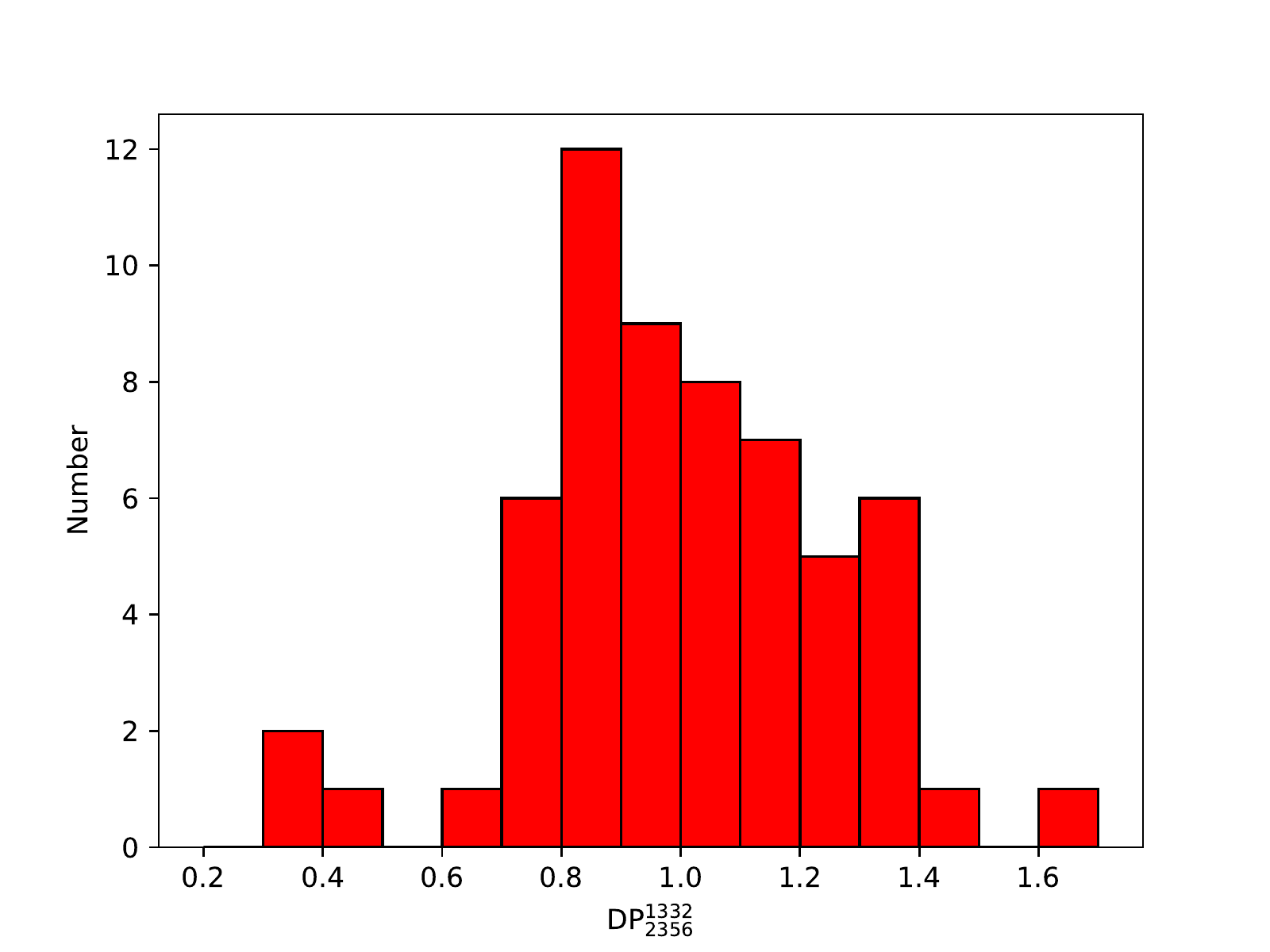}
\caption{Distributions of the total intensity spectral index (top), and polarised intensity spectral index (middle), and depolarisation (bottom) of the XXL$_{\text{39}}$ dataset.}
\label{fig:XXL39Indices}
\end{figure}

The spectral indices of this population were calculated by performing a least-squares linear fit in the log-log plane to our three lower frequency wavebands (1332, 1844, and 2356 MHz) as our sources were generally difficult to identify reliably in our 2868 MHz mosaic. We derived the error using a Monte Carlo method, repeatedly varying the flux density within errors at each waveband and refitting them, taking 
the standard deviation of this distribution as the error. The total intensity and polarised intensity spectral indices ($\alpha_S$ and $\alpha_P$, with the convention $S_{\nu}\propto\nu^{-\alpha}$) are shown in Table~\ref{tab:SourceList}, while the distributions of the indices are shown in Figure~\ref{fig:XXL39Indices}. In the case of multi-component sources, we calculated the spectral index for each individual component and the source as a whole.

We derive the mean total intensity spectral index of 0.62 $\pm$ 0.04 with a standard deviation of 0.31 and the mean polarised intensity spectral index of 0.55 $\pm$ 0.07 with a standard deviation of 0.53. The distributions, shown in Figure \ref{fig:XXL39Indices}, are statistically similar. Of particular note, however, are some cases where the polarised spectral index is inverted while the total intensity spectral index behaves normally. This may indicate particularly high levels of Faraday depolarisation towards the source.

In order to estimate the level of depolarisation of the sources, we derive DP$^{1332}_{2356} = \left(\frac{1332}{2356}\right)^{\alpha_S-\alpha_P}$, as an estimate of the ratio of fractional polarisations ($p_\nu = P_\nu / S_\nu \approx \nu^{(\alpha_P- \alpha_S)}$)
at these two frequencies, $p_{1332 {\rm MHz}}/p_{2356 {\rm MHz}}$. For standard Faraday depolarisation laws \citep[e.g.][]{Burn66,Sokoloff98}, DP$^{1332}_{2356} < 1$ (corresponding to a lower fractional polarisation at higher wavelengths). The distribution of DP$^{1332}_{2356}$, also shown in Figure \ref{fig:XXL39Indices}, has a mean of 1.00 $\pm$ 0.03 with a standard deviation of 0.26, indicating low levels of depolarisation and repolarisation across the population as a whole. For the sources with inverted $\alpha_P$, we find lower values of DP$^{1332}_{2356}$, consistent with the suspected high levels of depolarisation. A significant proportion ($\sim47$\%) of our population do exhibit repolarisation. This is not necessarily unexpected, for instance, \citet{Farnes14} found that 21\% of the polarised sources in their sample from the NRAO Very Large Array Survey (NVSS) exhibited repolarisation, which could be caused by differential Faraday rotation by a helical field \citep[e.g][]{Homan12,Horellou14} or by different emitting regions within the source. Several of our multi-component sources also exhibit significant differences in depolarisation between components, supporting \citeauthor{Farnes14}'s conclusion that the spectral energy distributions of polarised sources are particularly affected by different emitting regions within the source. Our overall finding is that the spectral indices generally match the results of other surveys at similar sensitivity levels and frequencies, indicating that the sources detected in the field represent a typical AGN population \citep[e.g.][]{Prandoni06,Grant10,Farnes14}.

Spectral indices can also give an indication of environment. From \citet{Bornancini10}, sources with $\alpha_S > 1$ are preferentially located within rich environments, such as clusters. This drops further to $\alpha_S > 0.65$ at low redshifts ($z \sim 0.2-0.3$). As shown in Table \ref{tab:SourceCluster}, few of our sources appear to be located in rich environments, which is consistent with the distribution of $\alpha_S$. The relatively flat spectra of the sources with unknown redshift plausibly indicates that they are at different, likely higher, redshifts than their corresponding cluster. Depolarisation can also give an indication of environment along the line of sight. In this case, however, the sources coincident with clusters exhibit comparable scatter from DP$^{1332}_{2356}=1$, $\sim 0.14$, with the general population, $\sim 0.20$. The greater scatter in the general population is dominated by a small number of highly depolarised or repolarised outliers, but it appears that the overall effects of the foreground cluster magnetic fields are small. Most of the sources we associate with clusters are located at $r\sim r_{200}$ for their respective clusters. Hence, both the distance travelled through the clusters' magnetic fields and the relative strength of those fields are significantly reduced.

\begin{table*}
\caption{Sources in the XXL$_{39}$ dataset that appear colocated with clusters from the catalogues detailed in \citetalias{Pacaud16}, \citetalias{Giles16} and \citetalias{Adami18}, the source and cluster redshifts (if available and favouring the spectroscopic redshifts from \citetalias{Chiappetti18}), total intensity and polarised intensity spectral indices ($\alpha _S$ and $\alpha _P$ where $S_{\nu} \propto \nu^{-\alpha}$), depolarisation as defined in Section \ref{sec:SpecIndex} (DP$^{1332}_{2356} = \left(\frac{1332}{2356}\right)^{\alpha_S-\alpha_P}$) and rotation measures (RMs). Means for the XXL$_{39}$ dataset are also included.}            % title of Table
\label{tab:SourceCluster}      % is used to refer this table in the text
\centering                          % used for centering table
\begin{tabular}{lccccccc}        % centered columns (3 columns)
\hline\hline                 % inserts double horizontal lines
Source & Source & Coincident & Cluster & $\alpha_S$ & $\alpha_P$ & DP$^{1332}_{2356}$  & RM \\
& redshift & cluster & redshift &&&& rad m$^{-2}$\\    % table heading 
\hline                        % inserts single horizontal line
   PKS 2319-55 & 0.878 & XLSSC 523 & 0.342 & 0.01 $\pm$ 0.02  & -0.13 $\pm$ 0.02 & 0.92 $\pm$ 0.01 & 14.3 $\pm$ 0.3 \\      % inserting body of the table
   SCSO J232419.6-552548.9 & 0.240 & XLSSC 612 & 0.275 & 0.34 $\pm$ 0.02  & 0.00 $\pm$ 0.73  & 0.82 $\pm$ 0.43 & 5.8 $\pm$ 7.9  \\
   SSTSL2 J232420.72-552532.0 & --- & XLSSC 612 & 0.275 & 0.89 $\pm$ 0.05  & 0.95 $\pm$ 0.32  & 1.03 $\pm$ 0.15 & 16.6 $\pm$ 4.7 \\
   SSTSL2 J232918.05-533902.7 & --- & XLSSC 622 & 0.276 & 0.26 $\pm$ 0.03  & 0.49 $\pm$ 0.14  & 1.14 $\pm$ 0.07 & ---   \\
   \quad Component A & --- & '' & '' & 0.44 $\pm$ 0.02  & 0.96 $\pm$ 0.14  & 1.35 $\pm$ 0.07 & 14.4 $\pm$ 6.9 \\
   \quad Component B & --- & '' & '' & -0.34 $\pm$ 0.02 & 0.15 $\pm$ 0.12  & 1.32 $\pm$ 0.06 & 13.2 $\pm$ 4.7 \\
   \quad Component C & --- & '' & '' & 0.47 $\pm$ 0.03  & 0.37 $\pm$ 0.17  & 0.94 $\pm$ 0.07 & 10.7 $\pm$ 9.3 \\
   WISEA J233035.37-533122.5 & 0.171 & XLSSC 623 & 0.171 & 0.04 $\pm$ 0.03  & 0.10 $\pm$ 1.14  & 1.04 $\pm$ 0.97 & 19.8 $\pm$ 1.4 \\
   2MASS J23320704-5444040 & 0.273 & XLSSC 524 & 0.270 & 0.77 $\pm$ 0.05  & 0.76 $\pm$ 0.03  & 0.99 $\pm$ 0.02 & ---   \\
   \quad Component A & '' & '' & '' & 0.84 $\pm$ 0.02  & 0.50 $\pm$ 0.03  & 0.82 $\pm$ 0.01 & 7.4 $\pm$ 3.1 \\
   \quad Component B & '' & '' & '' & 1.00 $\pm$ 0.02  & 0.94 $\pm$ 0.05  & 0.97 $\pm$ 0.02 & 16.0 $\pm$ 4.4 \\
   \quad Component C & '' & '' & '' & 0.52 $\pm$ 0.02  & 0.89 $\pm$ 0.08  & 1.23 $\pm$ 0.04 & 16.8 $\pm$ 4.2 \\
\hline
XXL$_{39}$ mean & --- & --- & --- & 0.62 $\pm$ 0.04 & 0.55 $\pm$ 0.07 & 1.00 $\pm$ 0.03 & 12.1 $\pm$ 0.9 \\
\hline                                   %inserts single line
\end{tabular}
\end{table*}

\subsection{Rotation measures}

\begin{figure}
\includegraphics[width=\columnwidth]{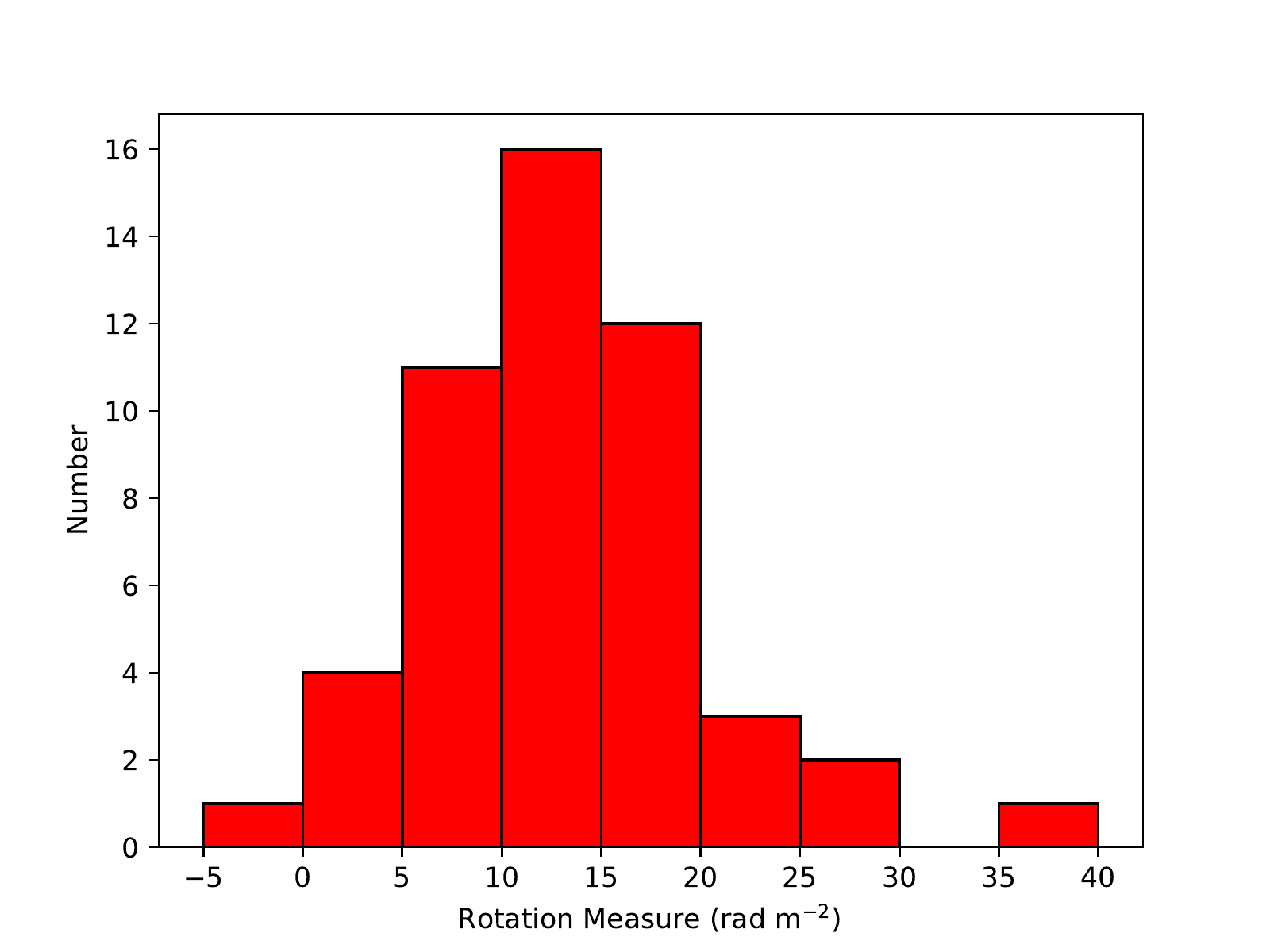}
\caption{Distribution of rotation measures of source components in the XXL$_{39}$ dataset.}
\label{fig:RMHist}
\end{figure}

Further measurements were performed on the $Q$ and $U$ maps at each waveband, using the source sizes and positions previously identified in order to gain a complete picture of the sources within the XXL$_{\text{39}}$ sample in all four frequency sub-bands (1332, 1844, 2356 and 2868 MHz). This also allowed the calculation of rotation measures for these sources, as presented in Table~\ref{tab:SourceList}. For sources consisting of several components, the rotation measures were calculated separately and no overall rotation measure for the source was derived as there were often large differences between the components. Our rotation measure distribution, as shown in Figure \ref{fig:RMHist}, has a mean of 12.1 $\pm$ 0.9 rad m$^{-2}$ with a standard deviation of 6.7 rad m$^{-2}$, which, as there are relatively few colocated sources and clusters, most likely indicates the approximate Galactic rotation measure in this field. It would be expected that the sources located towards clusters have a greater scatter due to additional contributions due to magnetic fields within the clusters \citep[e.g.][]{Kim91,Clarke01,Clarke04}. We find, however, that these sources, as shown in Table \ref{tab:SourceCluster}, have comparable deviations from 12.1 rad m$^{-2}$ as the general population with means of 3.9 and 5.3 rad m$^{-2,}$ respectively. This indicates that the additional rotation measures from the clusters are relatively weak, which is consistent with the fact that most of the sources are not located centrally behind a cluster and with our depolarisation findings.

\citet{Bernardi06} share 15 common sources of their catalogue of 18. The derived rotation measures were used to calculate the polarisation angles for our sources at the 1.4 GHz frequency examined by \citet{Bernardi06}. We find, however, that there is a large discrepancy, most likely due to \citet{Bernardi06} failing to fully account for the degeneracy of $\arctan(U/Q)$ as indicated by their calculated polarisation angles all being set between -45\degr and 45\degr.

\section{Leakage analysis}
\label{sec:Leakage}

Radio interferometers typically measure two orthogonal electric fields and, following corrections, use these to extract the source's Stokes parameters, $I$, $Q$, $U$ and $V$. The unintended response of the polarisation feed systems to the incorrect Stokes parameter, leakage, arises due to imperfections in the reflector and feed systems of the instrument \citep{Conway69}. In particular, the apparent change in the polarisation state of the instrument for off-axis sources can severely distort the polarisation of the sidelobes \citep{Morris64}.

The on-axis leakage terms (the so-called `D' terms) are readily measured by the conventional polarisation calibration, and are small for the ATCA\footnote{\url{https://www.atnf.csiro.au/observers/memos/AT39.9_129.pdf}}. However, off-axis effects will involve a correction function which depends on offset from the pointing centre. There are two main aspects of the widefield leakage pattern that can affect polarised observations: the off-axis separation and the relative direction of the source from the pointing centre.

Off-axis separation tends to be the dominant factor for the widefield leakage pattern, which can reach tens of percent \citep{Sutinjo15}. However, these extreme values are reached only for sources far off axis, and the leakage term generally only increases slowly with off-axis angle. In the case of the ATCA, previous investigations have indicated values of $\sim0.001-0.002$ polarisation percentage increase per arcsecond \citep{Sault99,Anderson15}.

Since our field was observed in multiple overlapping pointings, polarised sources appear in several different pointings. By comparing the individual pointing data for each source, it is possible to derive some characteristics of the widefield leakage pattern and the correction function. The 1332 MHz band was used for this analysis. The relative positions of the sources and pointings are shown in Figures~\ref{fig:IMosaic} and~\ref{fig:PMosaic}. Only the four closest pointings to each of the XXL$_{\text{39}}$ sources were used in this case.

\subsection{Linear off-axis separation}

The rate at which apparent polarisation percentage changes in line with separation from pointing centre was evaluated by performing a least-squares linear fit to the polarisation percentage measured in the different pointings against the source's separations from the pointing centres.

\begin{figure}
\centering
\includegraphics[width=\columnwidth]{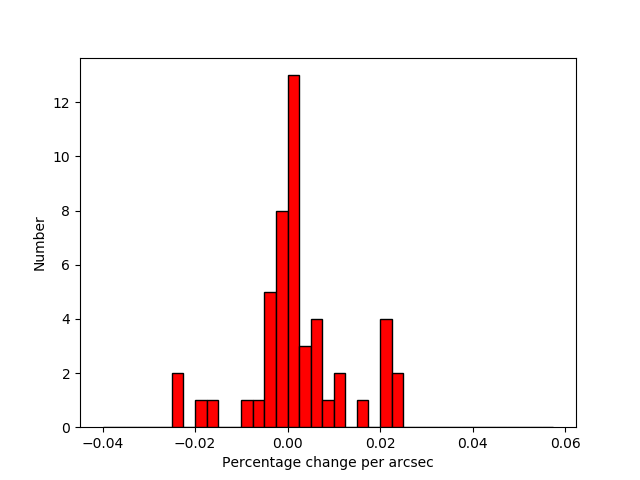}
\caption{Distribution of polarisation percentage change per arcsecond off-axis separation for the XXL$_{\text{39}}$ dataset.}
\label{fig:FullSampleHisto}
\end{figure}

The distribution of polarisation percentage change per arcsecond separation is shown in Figure~\ref{fig:FullSampleHisto}. The mean value was found to be 0.0023 $\pm$ 0.0015 percent per arcsecond, indicating, with a low significance, that leakage increases with angular separation from the pointing centre. This value is consistent with the values of $\sim0.0015$ percent per arcsecond found by \citet{Sault99} for their observations of Vela-X and $\sim0.0018$ percent per arcsecond found by \citet{Anderson15} for their observations of sources offset from the pointing centre by more than 0.155\degr.

The standard deviation of this distribution was found to be 0.0105, and the scatter and mean were dominated by a few sources lying further than 1.5 $\sigma$ from the mean. Part of this can be explained by fainter sources with greater errors in measurements for the $Q$ and $U$ which, in turn, leads to uncertainties in the change in polarisation percentage. These factors reduce the significance of the results. 

However, the data still indicate a mean total increase in polarisation percentage of $\sim$1.4\% at the locations which are most remote from the pointing grid centres. This is a relatively small increase when compared, for instance, to the variation of the leakage with frequency, and suggests that using a beam separation of $\lesssim$2/3 FWHM of the primary beam is an effective strategy for dealing with polarisation leakage in ATCA mosaics.

The dependence of the polarisation position angle on the off-axis separation was also plotted. The results indicate a slightly greater change in the polarisation position angle in sources with higher absolute polarisation position angles. This indicates higher leakage $Q$ as this effect is more dominant at higher angles than the $U$. However, several sources contradict the trend. This reflects the fact that the situation is more complex and leakage depends on multiple factors for alt-az mount telescopes such as the ATCA.

A small number of the brightest unpolarised sources were also examined to see if of-axis polarisation was introduced. It was found that these follow the same trends as polarised sources and they have a similar, albeit more significantly detected, mean change in  polarisation percentage per arcsec, 0.0014 $\pm$ 0.0005. This small change is again consistent with the results of both \citet{Sault99} and \citet{Anderson15}.

\subsection{Relative source-pointing position angle}

Another possible factor influencing polarisation leakage is the relative position angle between the pointing centre and the source. Again, both the polarisation percentage and how the polarisation position angle changes with relative position angle were investigated.

The polarisation percentage was plotted against these relative position angles. There are no obvious trends in the polarisation percentage and the distributions seem to be symmetrical around 0\degr, as expected. It appears, therefore, that the relative position angle of the source and pointing has little effect on the overall polarisation percentage of the source. However, the limited number of data points makes the conclusion tentative.

The polarisation position angle was then plotted against this pointing-source relative position angle for each source. This could indicate how the individual linear polarisations, $Q$ and $U$, are affected. Once again, there is no strong apparent trend and the distributions are symmetrical around 0\degr,\, as expected.

This analysis was repeated for a small number of bright, unpolarised sources. These indicated the expected symmetry around 0\degr\  but no other obvious trends.

\subsection{Discussion}

The results of this investigation indicate that polarisation leakage tends to increase with greater separation from the pointing centre and that $Q$ seems to be more affected by this phenomenon than $U$. This is in agreement with \citet{Sutinjo15} where their Figures 2 and 3 demonstrate the greater effect of leakage at greater zenith angles, that is, more off-axis in \citeauthor{Sutinjo15}'s configuration, and their finding that the absolute calculated $Q$ leakages are greater than that of $U$, as summarised in their Table 2.

Overall, however, this exploration of leakage in the ATCA is impacted by the lack of data. The result is based on four data points for each source which means that while our investigation indicates characteristics of the full correction function, we cannot currently derive it fully. There is also the possibility that any leakage terms are sufficiently small to have a negligible effect on polarised results, but this cannot be confirmed without additional data.

An ideal follow-up survey would look at fewer sources and observe each of them from many pointings at different separations and position angles relative to each source. The source population chosen for this survey should consist of several bright sources with known high polarisations.

\section{Conclusions}

We detected 39 polarised sources with a polarised flux density greater than 0.2 mJy in the central 5.634 deg$^2$ region of the XXL-South field. This polarised source count is similar to that previously reported by \citet{Rudnick14a}, although our apparent higher number density of sources at $\sim 1$ mJy compared to \citet{Grant10} might indicate a deficit particular to their observed fields or an excess particular to ours. The spectral indices and rotation measures were also examined and agree with expectations for a typical sample of AGN. Comparing our source population with the locations of X-ray identified clusters from \citetalias{Pacaud16}, \citetalias{Giles16} and \citetalias{Adami18} indicates that the depolarisation and rotation measure properties of sources towards clusters are consistent with those of the population as a whole and, therefore, the clusters' contributions to these properties are small.

By now, the entire XXL-South field has been fully observed at 2.1 GHz \citepalias{Butler18a} and expanding this study of the source population to allow a more in depth examination of the source properties should prove relatively simple. In addition, the XXL-South field is due to be covered by the Polarisation Sky Survey of the Universe's Magnetism (POSSUM\footnote{\url{http://askap.org/possum}}) with the Australia Square Kilometre Array Pathfinder (ASKAP). The greatly improved sensitivity of the 10 $\mu$Jy~beam$^{-1}$ at a resolution of 10\arcsec\, will yield a significant increase in the number of detected polarised sources.

The polarisation leakage effects of ATCA were also investigated in this paper. We find that leakage tends to increase at higher separations from the pointing centre as expected and that the $Q$ response seems to be more affected than the $U$. We find that leakage of less than 1.4\% of $I$ into $P$ occurs for the current mosaic, and a beam separation of $\lesssim$ 2/3 FWHM of the primary beam controls polarisation leakage at this level for the ATCA. The effect of the position angle of the source relative to the pointing centre was examined but no trend in either the polarisation percentage or polarisation position angle was observed. Mosaic observations of a richer field should allow a polarisation correction function to be derived for the ATCA.

\begin{acknowledgements}
    XXL is an international project based around an XMM Very Large Programme surveying two 25 \degsq extragalactic fields at a depth of $\sim 6\times 10^{15}$ erg cm$^{-2}$ s$^{-1}$ in the [0.5-2] keV band for point-like sources. The XXL website is \url{http://irfu.cea.fr/xxl}.

    Multi-band information and spectroscopic follow-up of the X-ray sources are obtained through a number of survey programmes, summarised at \url{http://xxlmultiwave.pbworks.com/}.

    The Australia Telescope Compact Array is funded by the Commonwealth of Australia for operation as a National Facility managed by CSIRO. 

    This research has made use of the NASA/IPAC Extragalactic Database (NED) which is operated by the Jet Propulsion Laboratory, California Institute of Technology, under contract with the National Aeronautics and Space Administration.

    RAJE acknowledges funding from the Science and Technology Facilities Council. RAJE would also like to thank the University of Bristol, particularly the School of Physics, for their invaluable support throughout this time. VS acknowledges funding from the European Union's Seventh Framework program under grant agreement 337595 (ERC Starting Grant, `CoSMass'). JD acknowledges financial assistance from the South African Radio Astronomy Observatory (SARAO; \url{www.ska.ac.za}).
    
    We thank L. Chiappetti for the technical report and G. Bernardi for useful comments and conversation. We also thank the referee for the useful comments.
\end{acknowledgements}

%-------------------------------------------------------------------

\bibliographystyle{aa}
\bibliography{XXL.bib}

\begin{figure*}
\includegraphics[width=\columnwidth]{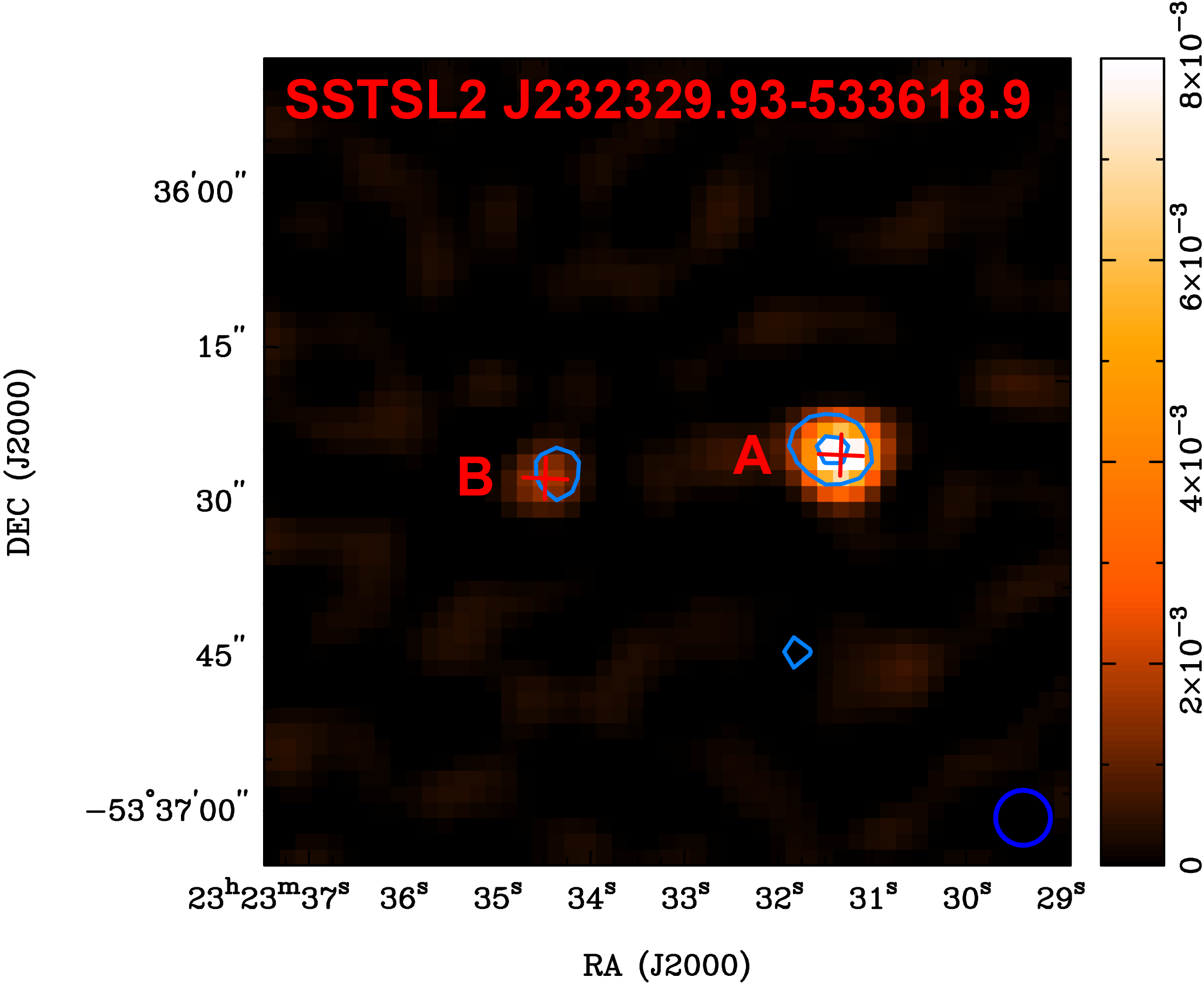}
\includegraphics[width=\columnwidth]{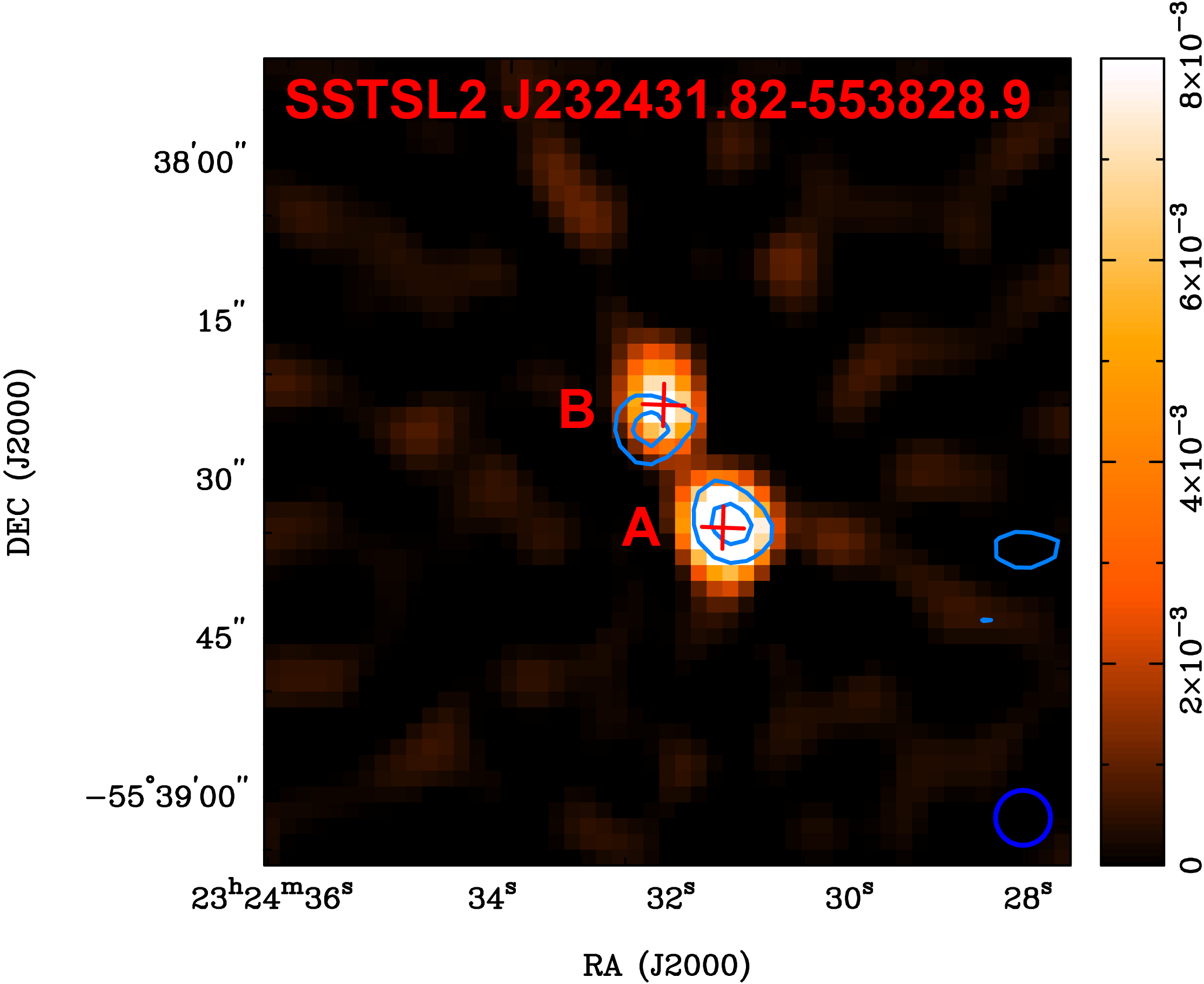}
\includegraphics[width=\columnwidth]{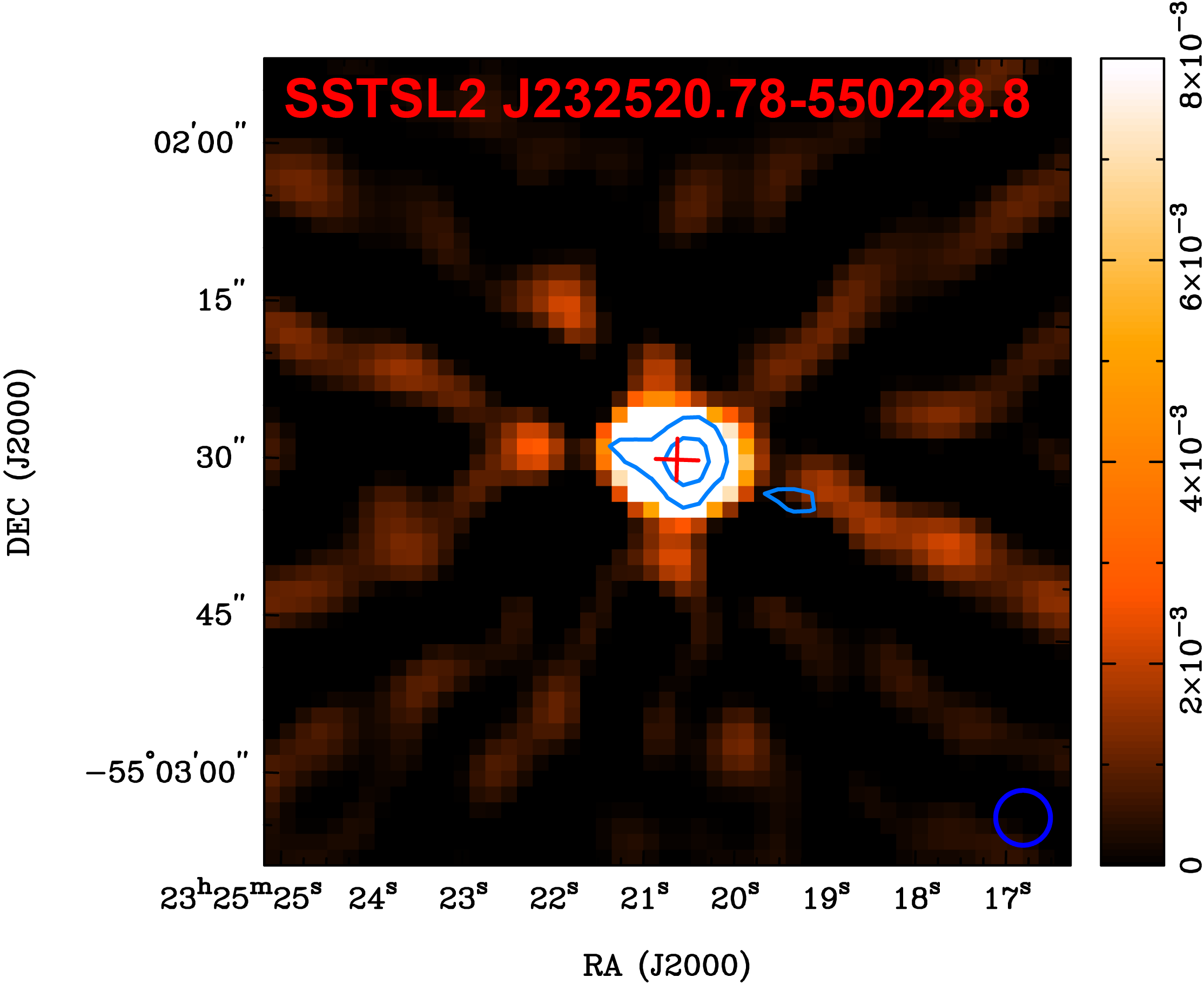}
\includegraphics[width=\columnwidth]{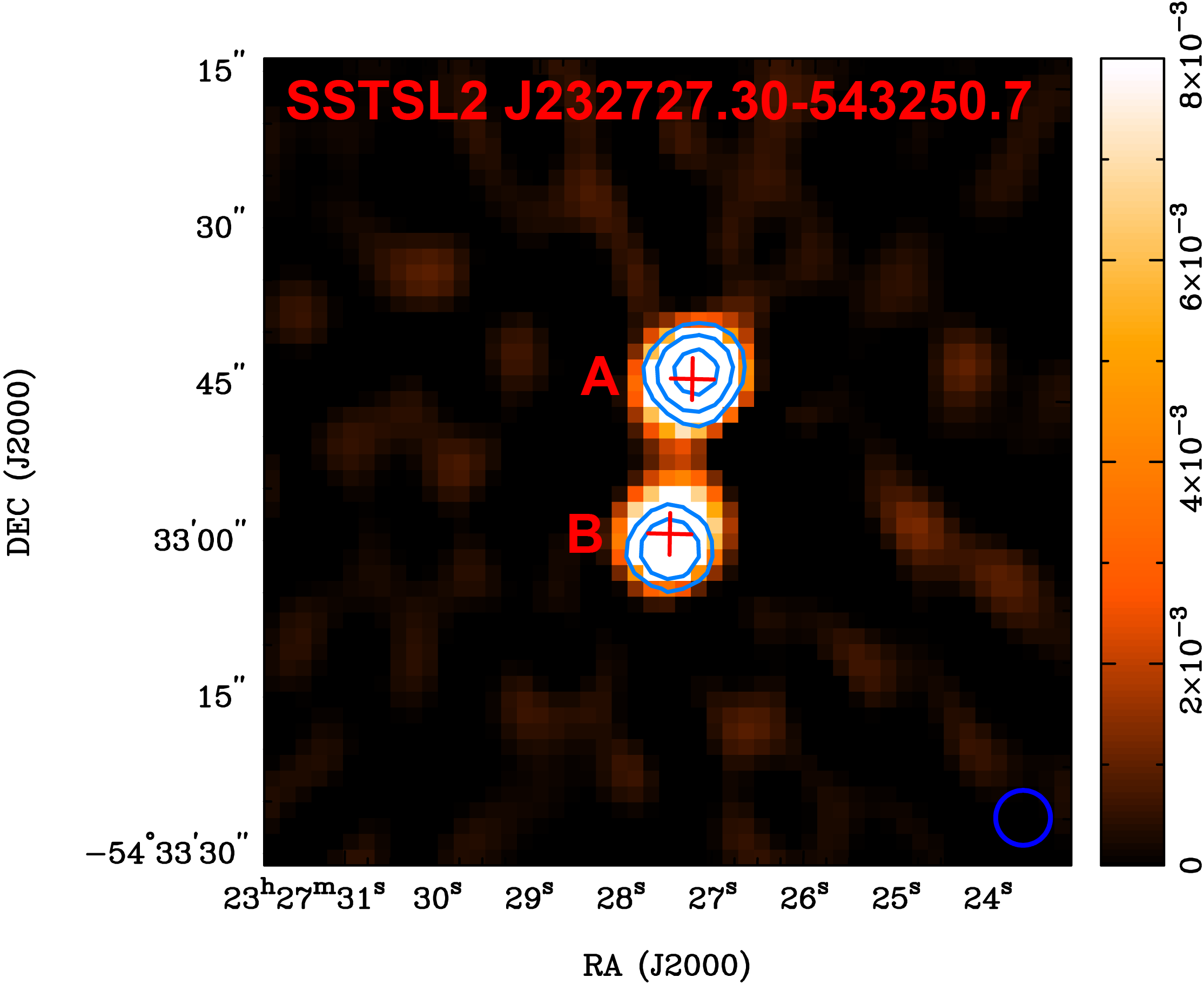}
\includegraphics[width=\columnwidth]{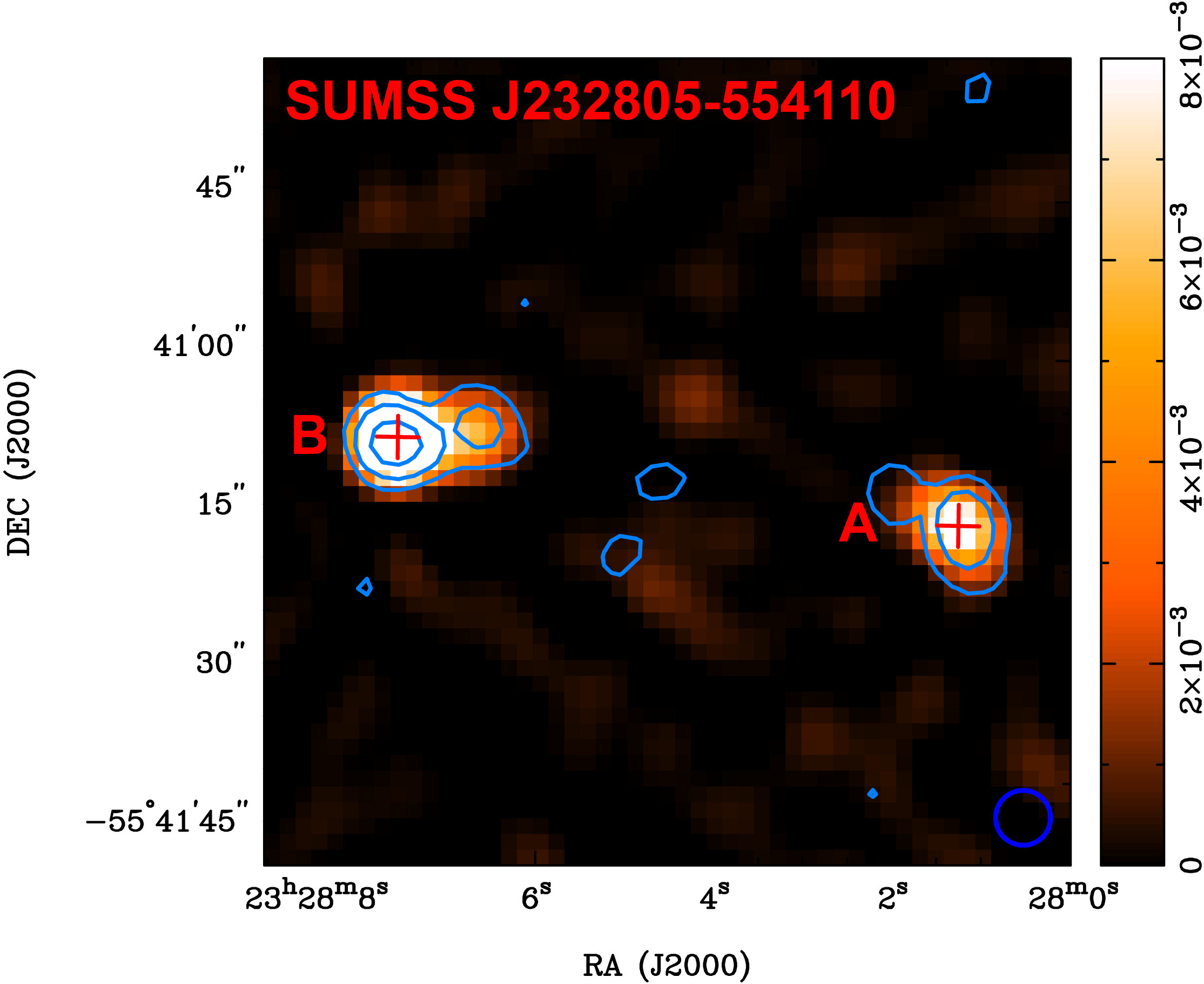}
\includegraphics[width=\columnwidth]{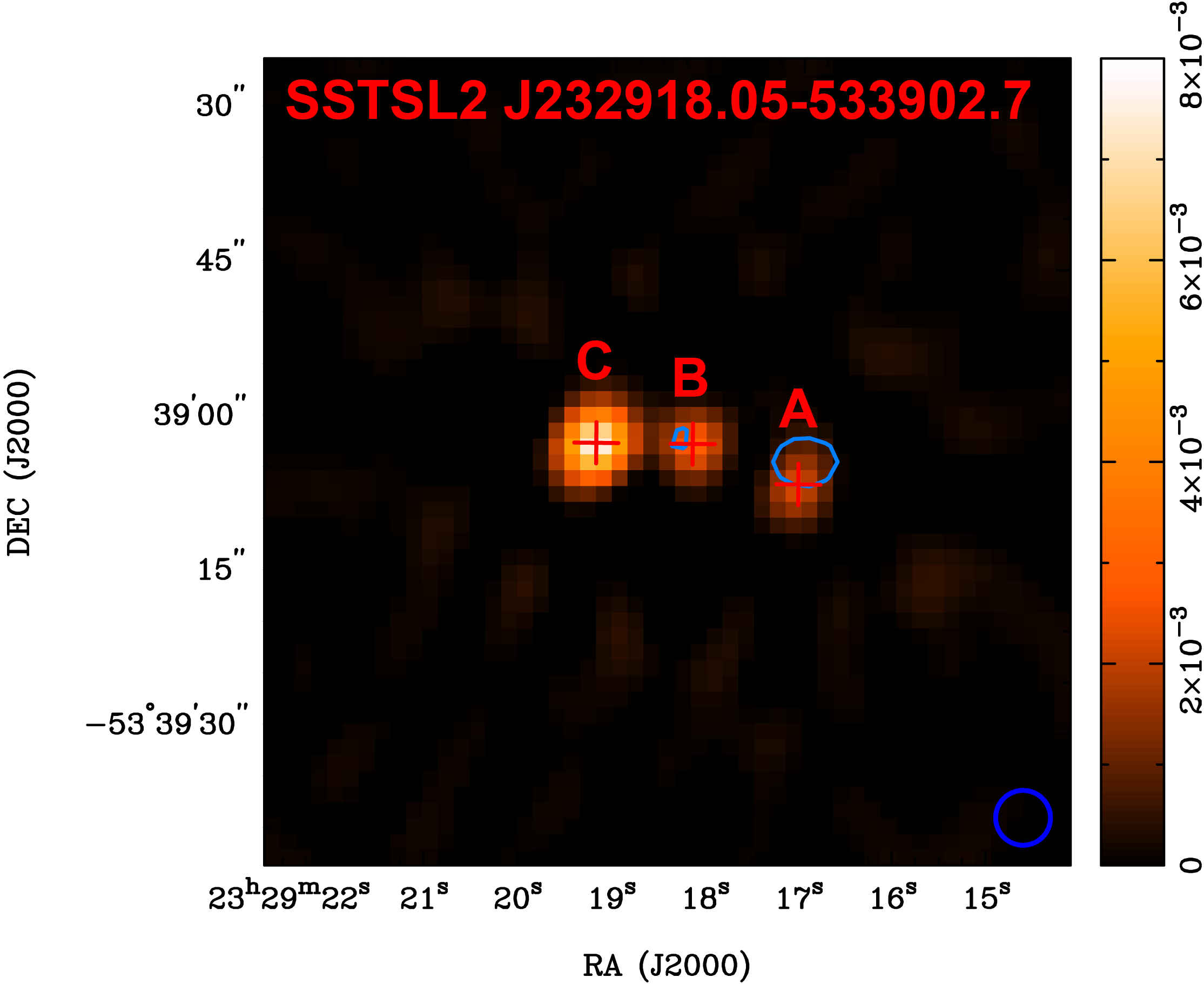}
\caption{Several example sources from the XXL$_{\text{39}}$ dataset, including all multi-component sources. Heat map indicates total intensity in Jy~beam$^{-1}$, azure contours indicate polarised intensity at 0.5, 1 and 2 mJy, and red crosses indicate the centre of each source component as identified by our fits to the total intensity. In the case of multi-component sources, each component is also labelled. The FWHM of the synthesised beam is shown in the bottom right corner of each plot. Continued in Appendix \ref{sec:appendix}.}
\label{fig:SourceExamples}
\end{figure*}

\appendix

\clearpage
\onecolumn

\begin{landscape}
\section{The XXL$_{39}$ dataset}
\label{sec:appendix}
\small
\begin{longtable}{lllllllllll}
\caption{The sources in the XXL$_{\text{39}}$ dataset. Shown in the table are each source's IAU designation, position, integrated total flux at 1332 MHz (S$_{1332}$), integrated polarised flux at 1332 MHz (P$_{1332}$), polarisation percentage, position angle ($\theta _{\text{PA}}$), total intensity and polarised intensity spectral indices ($\alpha _S$ and $\alpha _P$ where $S_{\nu} \propto \nu^{-\alpha}$), depolarisation as defined in Section \ref{sec:SpecIndex} (DP$^{1332}_{2356} = \left(\frac{1332}{2356}\right)^{\alpha_S-\alpha_P}$) and rotation measures (RMs). The positions given were measured by us as described in Section \ref{sec:SourceID} and are independent of those described in \citetalias{Butler18a}. We note that some of these sources have been split into multiple components for analysis. The properties of the combined source and each component are displayed in separate rows.}
\label{tab:SourceList} \\
\hline\hline
IAU Designation   & RA  & Dec   & S$_{1332}$  & P$_{1332}$  & Percentage  & $\theta_{\text{PA}}$ & $\alpha_S$    & $\alpha_P$ & DP$^{1332}_{2356}$  & RM  \\
 & J2000       & J2000 & mJy    & mJy   & polarised    & degrees   &       &   &    & r   ad m$^{-2}$  \\
\hline
\endfirsthead
\caption{continued.}\\
\hline\hline
IAU Designation   & RA  & Dec   & S$_{1332}$  & P$_{1332}$  & Percentage  & $\theta_{\text{PA}}$ & $\alpha_S$    & $\alpha_P$   & DP$^{1332}_{2356}$ & RM  \\
 & J2000       & J2000 & mJy    & mJy   & polarised    & degrees   &       &   &    & rad m$^{-2}$   \\
\hline
\endhead
\hline
\endfoot
PKS 2319-55 & 350.528750 & -54.757861  & 235.30 $\pm$ 9.52 & 67.94 $\pm$ 1.09 & 28.9 $\pm$ 1.3 & 31 $\pm$ 5 & 0.01 $\pm$ 0.02  & -0.13 $\pm$ 0.02 & 0.92 $\pm$ 0.01 & 14.3 $\pm$ 0.3   \\
SUMSS J232326-551331 & 350.857800 & -55.225263 & 21.52 $\pm$ 0.30 & 1.67 $\pm$ 0.26 & 7.7 $\pm$ 1.3 & 12 $\pm$ 3 & 1.36 $\pm$ 0.02 & 1.10 $\pm$ 0.05 & 0.86 $\pm$ 0.03 & 17.0 $\pm$ 6.4 \\
SSTSL2 J232329.93-533618.9   & 350.884667 & -53.606667  & 13.13 $\pm$ 0.66  & 2.12 $\pm$ 0.37  & 16.2 $\pm$ 2.9     & -5 $\pm$ 13   & 0.61 $\pm$ 0.02  & 0.90 $\pm$ 0.14  & 1.18 $\pm$ 0.07 & ---     \\
\quad Component A & 350.881083 & -53.606417  & 11.32 $\pm$ 0.43  & 1.44 $\pm$ 0.16  & 12.7 $\pm$ 1.5    & -85 $\pm$ 33  & 0.53 $\pm$ 0.02  & 0.86 $\pm$ 0.17 & 1.21 $\pm$ 0.08 & 5.8 $\pm$ 7.3    \\
\quad Component B & 350.894167 & -53.607389  & 1.80 $\pm$ 0.22   & 0.68 $\pm$ 0.20  & 37.8 $\pm$ 12.2  & -60 $\pm$ 41  & 1.14 $\pm$ 0.05  & 0.92 $\pm$ 0.14  & 0.88 $\pm$ 0.06 & 27.0 $\pm$ 18.9   \\
MRSS 191-011762 & 350.912750 & -53.640667  & 18.35 $\pm$ 0.64  & 2.10 $\pm$ 0.51  & 11.5 $\pm$ 2.8    & -52 $\pm$ 35  & 0.52 $\pm$ 0.02  & 0.31 $\pm$ 0.14  & 0.89 $\pm$ 0.05 & 3.0 $\pm$ 2.3    \\
SSTSL2 J232346.32-533847.1 & 350.943083 & -53.646611  & 29.39 $\pm$ 0.44  & 1.13 $\pm$ 0.17  & 3.8 $\pm$ 0.6      & 18 $\pm$ 33   & 0.47 $\pm$ 0.05  & 0.22 $\pm$ 0.14  & 0.87 $\pm$ 0.05 & 17.6 $\pm$ 4.8 \\
SUMSS J232355-534809 & 350.979250 & -53.805500  & 4.44 $\pm$ 0.43   & 2.27 $\pm$ 0.11  & 51.2 $\pm$ 5.5     & 63 $\pm$ 8    & 1.45 $\pm$ 0.05  & 1.21 $\pm$ 0.07  & 0.87 $\pm$ 0.04 & 11.1 $\pm$ 2.5   \\
SCSO J232419.6-552548.9      & 351.081625 & -55.430111  & 23.15 $\pm$ 1.88  & 0.59 $\pm$ 0.62  & 2.5 $\pm$ 2.7      & 8 $\pm$ 18    & 0.34 $\pm$ 0.02  & 0.00 $\pm$ 0.73  & 0.82 $\pm$ 0.43 & 5.8 $\pm$ 7.9  \\
SSTSL2 J232420.72-552532.0   & 351.084875 & -55.426028  & 9.40 $\pm$ 0.76   & 1.38 $\pm$ 0.70  & 14.7 $\pm$ 7.5     & 11 $\pm$ 8    & 0.89 $\pm$ 0.05  & 0.95 $\pm$ 0.32  & 1.03 $\pm$ 0.15 & 16.6 $\pm$ 4.7 \\
SSTSL2 J232431.82-553828.9   & 351.133292 & -55.641000  & 40.15 $\pm$ 3.60  & 3.20 $\pm$ 1.20  & 8.0 $\pm$ 3.1      & -72 $\pm$ 15  & 1.02 $\pm$ 0.03  & 0.60 $\pm$ 0.22  & 0.79 $\pm$ 0.08 & ---     \\
\quad Component A & 351.131125 & -55.642472 & 27.80 $\pm$ 1.80  & 1.83 $\pm$ 0.60  & 6.6 $\pm$ 2.2      & 20 $\pm$ 23   & 1.05 $\pm$ 0.02  & 0.69 $\pm$ 0.19  & 0.81 $\pm$ 0.06 & 9.7 $\pm$ 3.7   \\
\quad Component B & 351.134208 & -55.639306 & 12.35 $\pm$ 1.80  & 1.37 $\pm$ 0.60  & 11.1 $\pm$ 5.1     & 19 $\pm$ 27   & 0.96 $\pm$ 0.03  & 0.48 $\pm$ 0.29  & 0.76 $\pm$ 0.09 & 19.1 $\pm$ 5.6  \\
SSTSL2 J232520.78-550228.8   & 351.336375 & -55.041361  & 106.80 $\pm$ 1.70 & 1.89 $\pm$ 0.57  & 1.8 $\pm$ 0.5      & 74 $\pm$ 3    & 0.58 $\pm$ 0.05  & 1.10 $\pm$ 0.19  & 1.35 $\pm$ 0.11 & 21.5 $\pm$ 3.1 \\
SUMSS J232614-540321 & 351.561000 & -54.058778  & 21.22 $\pm$ 0.58  & 2.07 $\pm$ 0.46  & 9.7 $\pm$ 2.2     & -3 $\pm$ 7    & 0.90 $\pm$ 0.02  & 1.44 $\pm$ 0.17  & 1.36 $\pm$ 0.09 & 12.1 $\pm$ 12.1   \\
SSTSL2 J232717.97-550936.2   & 351.824708 & -55.160111  & 84.45 $\pm$ 0.54  & 1.64 $\pm$ 0.47  & 1.9 $\pm$ 0.6     & -15 $\pm$ 11  & 0.72 $\pm$ 0.05  & 0.69 $\pm$ 0.18  & 0.98 $\pm$ 0.18 & 27.6 $\pm$ 3.2 \\
SSTSL2 J232724.36-545111.0   & 351.852708 & -54.853889 & 81.09 $\pm$ 1.84  & 3.15 $\pm$ 0.62  & 3.9 $\pm$ 0.8      & -49 $\pm$ 12  & 0.86 $\pm$ 0.05  & -0.66 $\pm$ 0.07 & 0.42 $\pm$ 0.01 & -3.9 $\pm$ 1.1  \\
SSTSL2 J232727.30-543250.7   & 351.864167 & -54.547528  & 98.44 $\pm$ 3.46  & 5.41 $\pm$ 0.37  & 5.5 $\pm$ 0.4      & -33 $\pm$ 9   & 0.84 $\pm$ 0.02  & 0.74 $\pm$ 0.09 & 0.94 $\pm$ 0.03    & ---     \\
\quad Component A & 351.863708 & -54.545444 & 59.73 $\pm$ 1.76  & 3.33 $\pm$ 0.19  & 5.6 $\pm$ 0.4      & -38 $\pm$ 6   & 0.80 $\pm$ 0.02  & 0.90 $\pm$ 0.07  & 1.06 $\pm$ 0.03 & 16.5 $\pm$ 1.4   \\
\quad Component B & 351.864625 & -54.549583 & 38.71 $\pm$ 1.71  & 2.08 $\pm$ 0.18  & 5.4 $\pm$ 0.5      & -29 $\pm$ 5   & 0.89 $\pm$ 0.02  & 0.48 $\pm$ 0.08  & 0.79 $\pm$ 0.03 & 11.3 $\pm$ 2.4   \\
SUMSS J232805-554110 & 352.017583 & -55.686278  & 53.30 $\pm$ 1.02  & 6.85 $\pm$ 0.87  & 12.9 $\pm$ 1.6     & -81 $\pm$ 13  & 0.60 $\pm$ 0.05  & 0.29 $\pm$ 0.08  & 0.84 $\pm$ 0.03 & ---     \\
\quad Component A & 352.005458 & -55.687722  & 13.44 $\pm$ 0.51  & 2.75 $\pm$ 0.43  & 20.5 $\pm$ 3.3     & 45 $\pm$ 5   & 0.91 $\pm$ 0.02  & 0.96 $\pm$ 0.10  & 1.03 $\pm$ 0.04 & 11.3 $\pm$ 3.8   \\
\quad Component B & 352.031542 & -55.685667   & 39.86 $\pm$ 0.51  & 4.10 $\pm$ 0.44  & 10.3 $\pm$ 1.1     & -90 $\pm$ 17  & 0.50 $\pm$ 0.05  & -0.03 $\pm$ 0.05 & 0.74 $\pm$ 0.03 & 0.7 $\pm$ 1.5    \\
SUMSS J232825-551508 & 352.104500 & -55.252389   & 96.79 $\pm$ 1.85  & 2.33 $\pm$ 0.62  & 2.4 $\pm$ 0.6      & -36 $\pm$ 8   & 0.58 $\pm$ 0.05  & -1.32 $\pm$ 0.07 & 0.39 $\pm$ 0.03 & 20.6 $\pm$ 1.3   \\
SSTSL2 J232835.15-544124.2   & 352.146833 & -54.690028  & 51.84 $\pm$ 0.98  & 5.97 $\pm$ 0.10  & 11.5 $\pm$ 0.3     & 75 $\pm$ 3    & 0.42 $\pm$ 0.05  & 0.24 $\pm$ 0.02  & 0.90 $\pm$ 0.02 & 14.7 $\pm$ 0.4 \\
SUMSS J232903-553319 & 352.261958 & -55.554722  & 5.94 $\pm$ 0.43   & 0.99 $\pm$ 0.35  & 16.7 $\pm$ 7.7    & -62 $\pm$ 12  & 0.68 $\pm$ 0.03  & 0.93 $\pm$ 0.25  & 1.15 $\pm$ 0.13 & 16.4 $\pm$ 7.1   \\
SSTSL2 J232918.05-533902.7   & 352.326000 & -53.651111  & 14.37 $\pm$ 1.20  & 1.87 $\pm$ 0.67  & 13.0  $\pm$ 4.8   & 94 $\pm$ 29   & 0.26 $\pm$ 0.03  & 0.49 $\pm$ 0.14  & 1.14 $\pm$ 0.07 & ---     \\
\quad Component A & 352.320958 & -53.651694  & 2.43 $\pm$ 0.23   & 0.74 $\pm$ 0.23  & 30.2 $\pm$ 9.7     & -15 $\pm$ 16  & 0.44 $\pm$ 0.02  & 0.96 $\pm$ 0.14  & 1.35 $\pm$ 0.07 & 14.4 $\pm$ 6.9   \\
\quad Component B & 352.325667 & -53.650639  & 3.08 $\pm$ 0.19   & 0.58 $\pm$ 0.21  & 19.0 $\pm$ 7.0     & 69 $\pm$ 87   & -0.34 $\pm$ 0.02 & 0.15 $\pm$ 0.12  & 1.32 $\pm$ 0.06 & 13.2 $\pm$ 4.7  \\
\quad Component C & 352.329958 & -53.650639  & 8.87 $\pm$ 0.78   & 0.55 $\pm$ 0.23  & 6.3  $\pm$ 2.7      & -45 $\pm$ 40  & 0.47 $\pm$ 0.03  & 0.37 $\pm$ 0.17  & 0.97 $\pm$ 0.04 & 10.7 $\pm$ 9.3   \\
SSTSL2 J232925.10-545435.7   & 352.354750 & -54.909639 & 176.70 $\pm$ 1.48 & 4.03 $\pm$ 0.50  & 2.3 $\pm$ 0.3     & -36 $\pm$ 1   & 0.64 $\pm$ 0.03  & 0.90 $\pm$ 0.07  & 1.16 $\pm$ 0.04 & 6.6 $\pm$ 1.0    \\
SUMSS J232942-542524 & 352.431417 & -54.422556  & 40.10 $\pm$ 0.56  & 5.81 $\pm$ 0.08  & 14.5 $\pm$ 0.3     & 56 $\pm$ 2    & 0.51 $\pm$ 0.03  & 0.64 $\pm$ 0.03  & 1.08 $\pm$ 0.02 & 10.4 $\pm$ 0.5   \\
SSTSL2 J233033.80-543146.6   & 352.640750 & -54.529722   & 6.11 $\pm$ 0.33   & 1.36 $\pm$ 0.28  & 22.3 $\pm$ 4.8     & 84 $\pm$ 24   & 0.54 $\pm$ 0.02  & 0.83 $\pm$ 0.10  & 1.18 $\pm$ 0.05 & ---     \\
\quad Component A & 352.627083 & -54.519111  & 2.60 $\pm$ 0.15   & 0.55 $\pm$ 0.12  & 21.3 $\pm$ 4.7     & 4 $\pm$ 11      & 0.58 $\pm$ 0.02  & 0.57 $\pm$ 0.10  & 0.99 $\pm$ 0.04 & 10.5 $\pm$ 45.4  \\
\quad Component B & 352.640083 & -54.530611  & 3.51 $\pm$ 0.18   & 0.81 $\pm$ 0.17  & 23.1 $\pm$ 4.9     & -78 $\pm$ 74  & 0.52 $\pm$ 0.02  & 1.02 $\pm$ 0.08  & 1.33 $\pm$ 0.05 & 11.0 $\pm$ 7.2   \\
WISEA J233035.37-533122.5      & 352.647208 & -53.522833 & 22.87 $\pm$ 0.44  & 0.54 $\pm$ 0.23  & 2.4 $\pm$ 0.1      & -61 $\pm$ 1   & 0.04 $\pm$ 0.03  & 0.10 $\pm$ 1.14  & 1.03 $\pm$ 0.97 & 19.8 $\pm$ 1.4  \\
SSTSL2 J233119.46-552702.3   & 352.831292 & -55.450583  & 12.76 $\pm$ 0.34  & 1.88 $\pm$ 0.28  & 14.8 $\pm$ 2.2     & 56 $\pm$ 2    & 0.59 $\pm$ 0.02  & 0.61 $\pm$ 0.08  & 1.01 $\pm$ 0.04 & 17.5 $\pm$ 2.1 \\
2MASS J23320704-5444040      & 353.029667 & -54.734444   & 8.75 $\pm$ 0.34   & 3.46 $\pm$ 0.43  & 39.6 $\pm$ 5.1     & 12 $\pm$ 7   & 0.77 $\pm$ 0.02  & 0.76 $\pm$ 0.05  & 0.99 $\pm$ 0.02 & ---     \\
\quad Component A & 353.026583 & -54.742306 & 2.79 $\pm$ 0.08   & 1.23 $\pm$ 0.09 & 44.2 $\pm$ 3,6    & 30 $\pm$ 6    & 0.84 $\pm$ 0.02  & 0.50 $\pm$ 0.03  & 0.82 $\pm$ 0.01 & 7.4 $\pm$ 3.1    \\
\quad Component B & 353.029750 & -54.733833  & 3.10 $\pm$ 0.09   & 1.28 $\pm$ 0.15  & 41.2 $\pm$ 5.2     & 11 $\pm$ 4    & 1.00 $\pm$ 0.02  & 0.94 $\pm$ 0.05  & 0.97 $\pm$ 0.02 & 16.0 $\pm$ 4.4   \\
\quad Component C & 353.030958 & -54.730694  & 2.86 $\pm$ 0.17   & 0.95 $\pm$ 0.18  & 33.2 $\pm$ 6.5     & 16 $\pm$ 8    & 0.52 $\pm$ 0.02  & 0.89 $\pm$ 0.08  & 1.23 $\pm$ 0.04 & 16.8 $\pm$ 4.2   \\
SUMSS J233253-551055 & 353.222542 & -55.179722  & 59.01 $\pm$ 1.39  & 8.63 $\pm$ 0.26  & 14.6 $\pm$ 0.9    & -21 $\pm$ 12  & 0.55 $\pm$ 0.03  & 0.80 $\pm$ 0.03  & 1.15 $\pm$ 0.03 & ---     \\
\quad Component A & 353.220000 & -55.184444  & 50.69 $\pm$ 1.01   & 6.52 $\pm$ 0.13  & 12.9 $\pm$ 0.5    & 13 $\pm$ 90   & 0.50 $\pm$ 0.02  & 0.66 $\pm$ 0.10  & 1.10 $\pm$ 0.04 & 35.3 $\pm$ 4.0   \\
\quad Component B & 353.225542 & -55.176194  & 8.32 $\pm$ 0.38   & 2.11 $\pm$ 0.10  & 24.7 $\pm$ 1.7     & 34 $\pm$ 17   & 0.91 $\pm$ 0.02  & 1.25 $\pm$ 0.08  & 1.21 $\pm$ 0.04 & 12.3 $\pm$ 2.6   \\
SSTSL2 J233311.94-541623.4   & 353.301792 & -54.273972  & 1.87 $\pm$ 0.40   & 1.52 $\pm$ 0.08  & 81.2 $\pm$ 17.9     & 59 $\pm$ 24   & 0.35 $\pm$ 0.08  & 0.72 $\pm$ 0.02  & 1.23 $\pm$ 0.04 & 8.5 $\pm$ 6.1 \\
SSTSL2 J233329.53-540940.6   & 353.373958 & -54.164167 & 2.74 $\pm$ 0.07   & 0.95 $\pm$ 0.10  & 34.7 $\pm$ 3.7    & -11 $\pm$ 7   & 0.73 $\pm$ 0.03  & 0.52 $\pm$ 0.05  & 0.89 $\pm$ 0.03 & 9.0 $\pm$ 3.7    \\
SSTSL2 J233354.39-545540.4   & 353.476125 & -54.927611  & 32.49 $\pm$ 0.26  & 1.88 $\pm$ 0.07  & 5.8 $\pm$ 0.2      & -67 $\pm$ 1   & 0.55 $\pm$ 0.03  & 0.26 $\pm$ 0.02  & 0.85 $\pm$ 0.01 & 10.0 $\pm$ 1.9 \\
SSTSL2 J233445.21-541907.8   & 353.687125 & -54.318722   & 18.53 $\pm$ 0.20  & 2.84 $\pm$ 0.10  & 15.3 $\pm$ 0.6     & 27 $\pm$ 29   & 0.59 $\pm$ 0.03  & 0.39 $\pm$ 0.02  & 0.89 $\pm$ 0.02 & 10.2 $\pm$ 0.9 \\
SSTSL2 J233500.11-545534.2   & 353.749917 & -54.926806  & 23.89 $\pm$ 0.32  & 0.92 $\pm$ 0.09  & 3.9 $\pm$ 0.4      & 47 $\pm$ 1    & 0.59 $\pm$ 0.03  & -0.08 $\pm$ 0.03 & 0.68 $\pm$ 0.02 & 16.7 $\pm$ 3.0 \\
SSTSL2 J233551.16-532227.5   & 353.963125 & -53.374000  & 69.72 $\pm$ 0.41  & 1.20 $\pm$ 0.26  & 1.7 $\pm$ 0.4      & 22 $\pm$ 2    & 0.81 $\pm$ 0.03  & -1.09 $\pm$ 0.08 & 0.34 $\pm$ 0.01 & 18.1 $\pm$ 2.3 \\
SSTSL2 J233619.37-550342.0   & 354.081333 & -55.061667  & 19.47 $\pm$ 0.37  & 2.68 $\pm$ 0.19  & 13.8 $\pm$ 1.0     & 7 $\pm$ 4     & 0.41 $\pm$ 0.03  & 1.08 $\pm$ 0.03  & 1.47 $\pm$ 0.03 & ---     \\
\quad Component A & 354.079500 & -55.063083 & 10.33 $\pm$ 0.19  & 1.72 $\pm$ 0.10  & 16.7 $\pm$ 1.0     & 3 $\pm$ 50    & 0.43 $\pm$ 0.03  & 0.98 $\pm$ 0.05  & 1.37 $\pm$ 0.03 & 11.3 $\pm$ 3.1   \\
\quad Component B & 354.081792 & -55.060222 & 9.14 $\pm$ 0.18   & 0.96 $\pm$ 0.09  & 10.5 $\pm$ 1.0     & 10 $\pm$ 49   & 0.39 $\pm$ 0.03  & 1.28 $\pm$ 0.05  & 1.66 $\pm$ 0.05 & 6.5 $\pm$ 8.3    \\
SSTSL2 J233632.91-532647.9   & 354.137500 & -53.446889  & 54.41 $\pm$ 0.33  & 4.28 $\pm$ 0.21  & 7.9 $\pm$ 0.4      & -37 $\pm$ 1   & 0.63 $\pm$ 0.03  & 0.77 $\pm$ 0.03  & 1.08 $\pm$ 0.02 & 5.3 $\pm$ 1.0  \\
SUMSS J233747-552711 & 354.450625 & -55.453306  & 10.89 $\pm$ 0.79  & 1.04 $\pm$ 0.79  & 9.6 $\pm$ 7.3      & 44 $\pm$ 8    & 0.62 $\pm$ 0.03  & 0.18 $\pm$ 0.25  & 0.78 $\pm$ 0.11 & 0.9 $\pm$ 4.6    \\
SSTSL2 J233838.02-545841.3   & 354.658208 & -54.978278  & 77.19 $\pm$ 0.15  & 5.24 $\pm$ 0.10  & 6.8 $\pm$ 0.1      & 16 $\pm$ 3    & 0.50 $\pm$ 0.03  & 0.69 $\pm$ 0.02  & 1.11 $\pm$ 0.02 & 5.1 $\pm$ 1.0  \\
MRSS 192-101439   & 354.728083 & -55.021667  & 18.33 $\pm$ 0.19  & 2.24 $\pm$ 0.10  & 12.2 $\pm$ 0.5     & 58 $\pm$ 22   & 0.42 $\pm$ 0.03  & -0.10 $\pm$ 0.02 & 0.74 $\pm$ 0.01 & 2.5 $\pm$ 1.9    \\
GALEXASC J233903.89-553358.2 & 354.766458 & -55.567194   & 16.95 $\pm$ 0.51   & 2.21 $\pm$ 0.15  & 13.1 $\pm$ 1.3   & -65 $\pm$ 28   & 0.25 $\pm$ 0.03  & 0.69 $\pm$ 0.02  & 1.29 $\pm$ 0.02 & 7.3 $\pm$ 2.2 \\
WISE J233913.22-552350.8     & 354.804917 & -55.397472 & 156.70 $\pm$ 0.81 & 3.70 $\pm$ 0.85  & 2.4 $\pm$ 0.5      & 40 $\pm$ 3    & 0.13 $\pm$ 0.03  & -0.04 $\pm$ 0.07  & 0.91 $\pm$ 0.03 & 23.6 $\pm$ 1.2   \\
\end{longtable}
\end{landscape}

\twocolumn
\renewcommand\thefigure{\arabic{figure}}
\setcounter{figure}{8}

\begin{figure*}[h]
\includegraphics[width=\columnwidth]{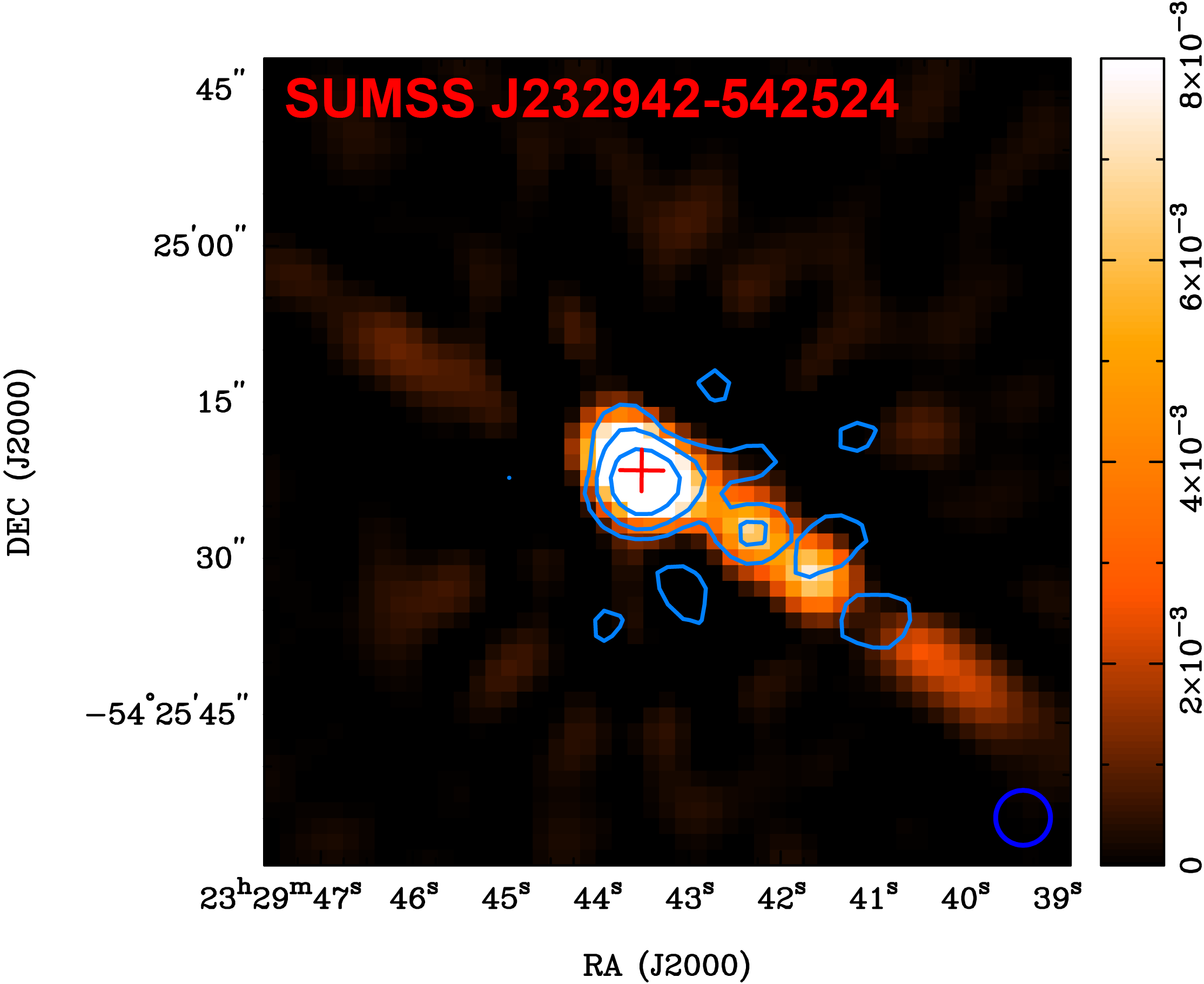}
\includegraphics[width=\columnwidth]{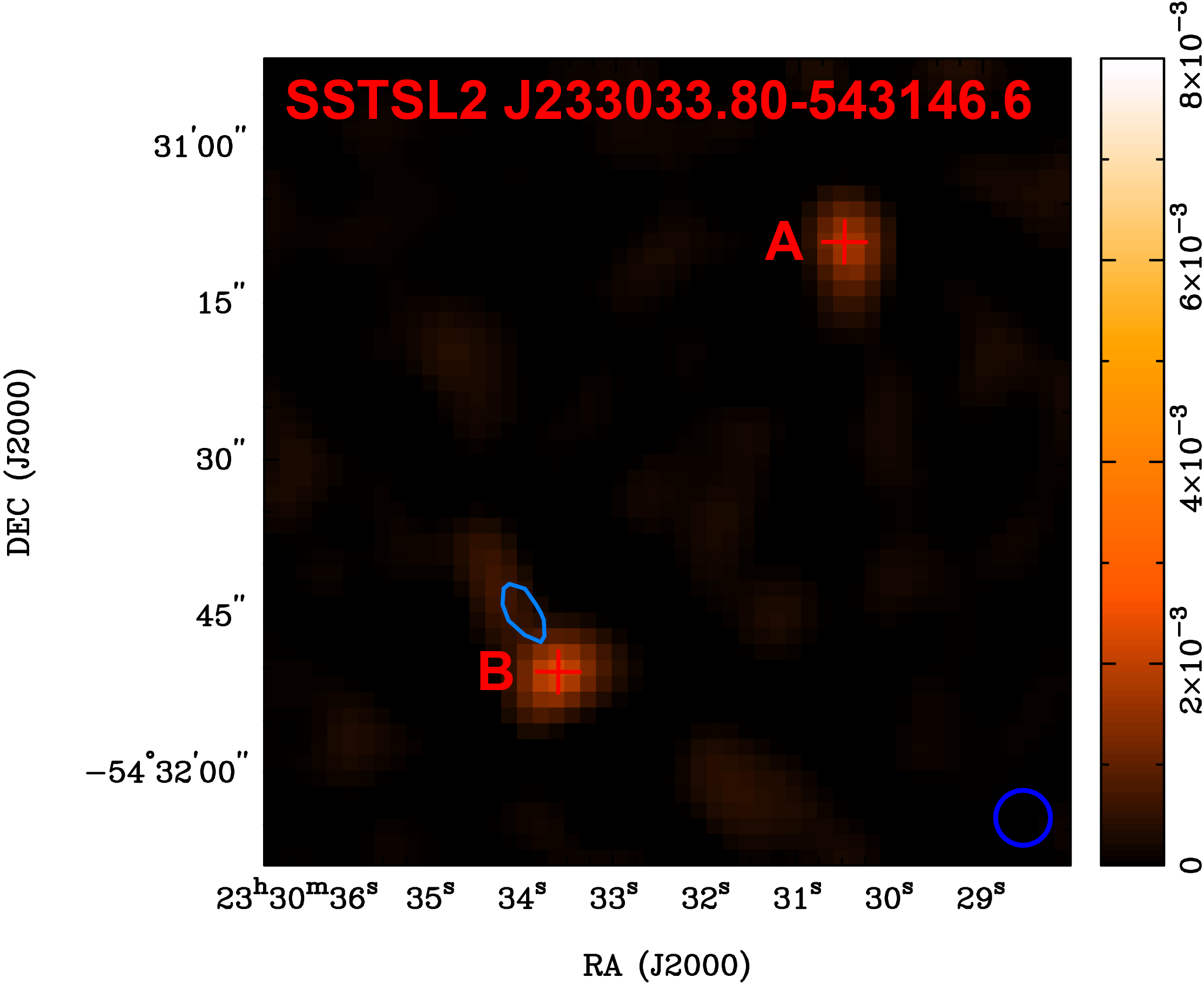}
\includegraphics[width=\columnwidth]{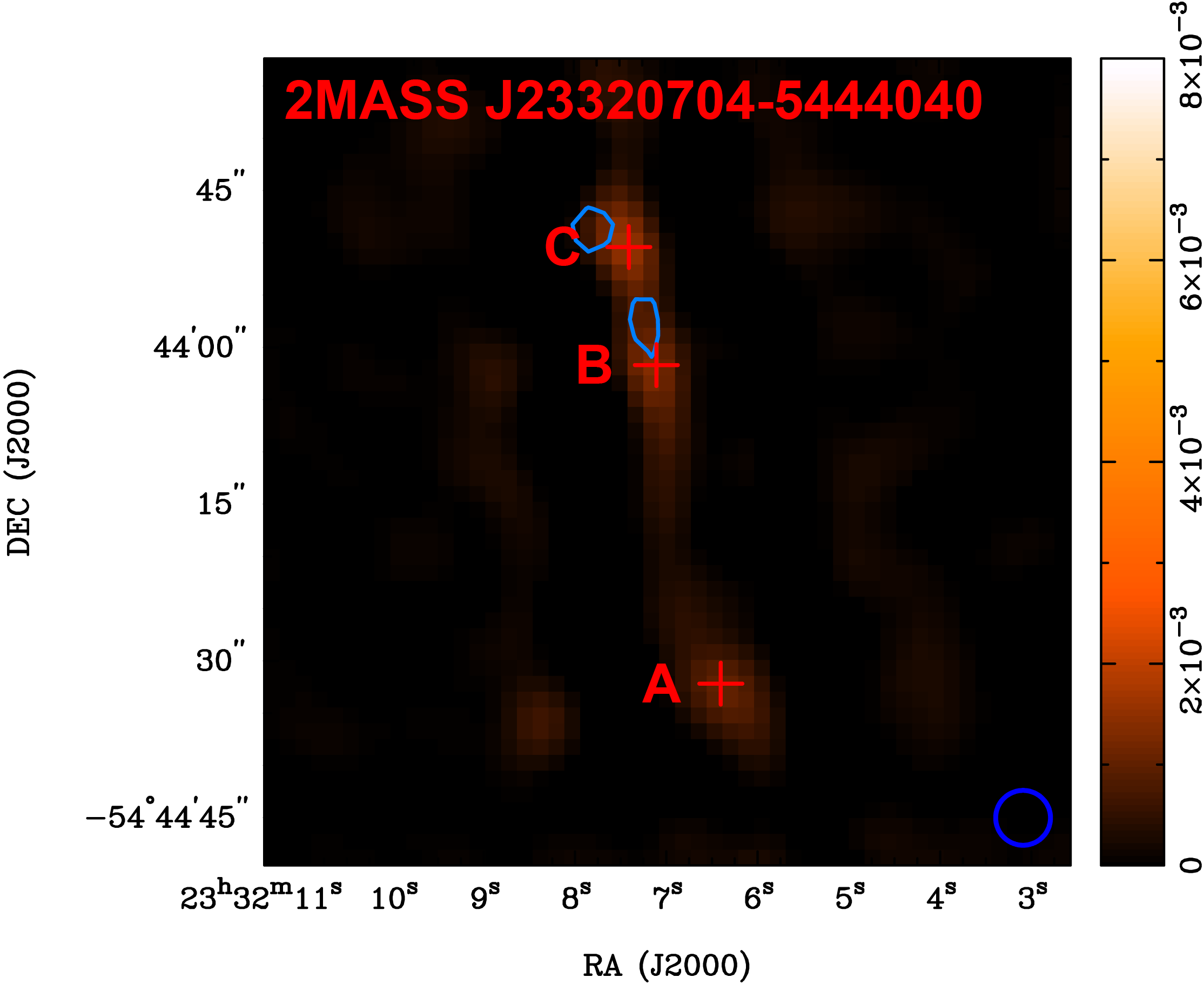}
\includegraphics[width=\columnwidth]{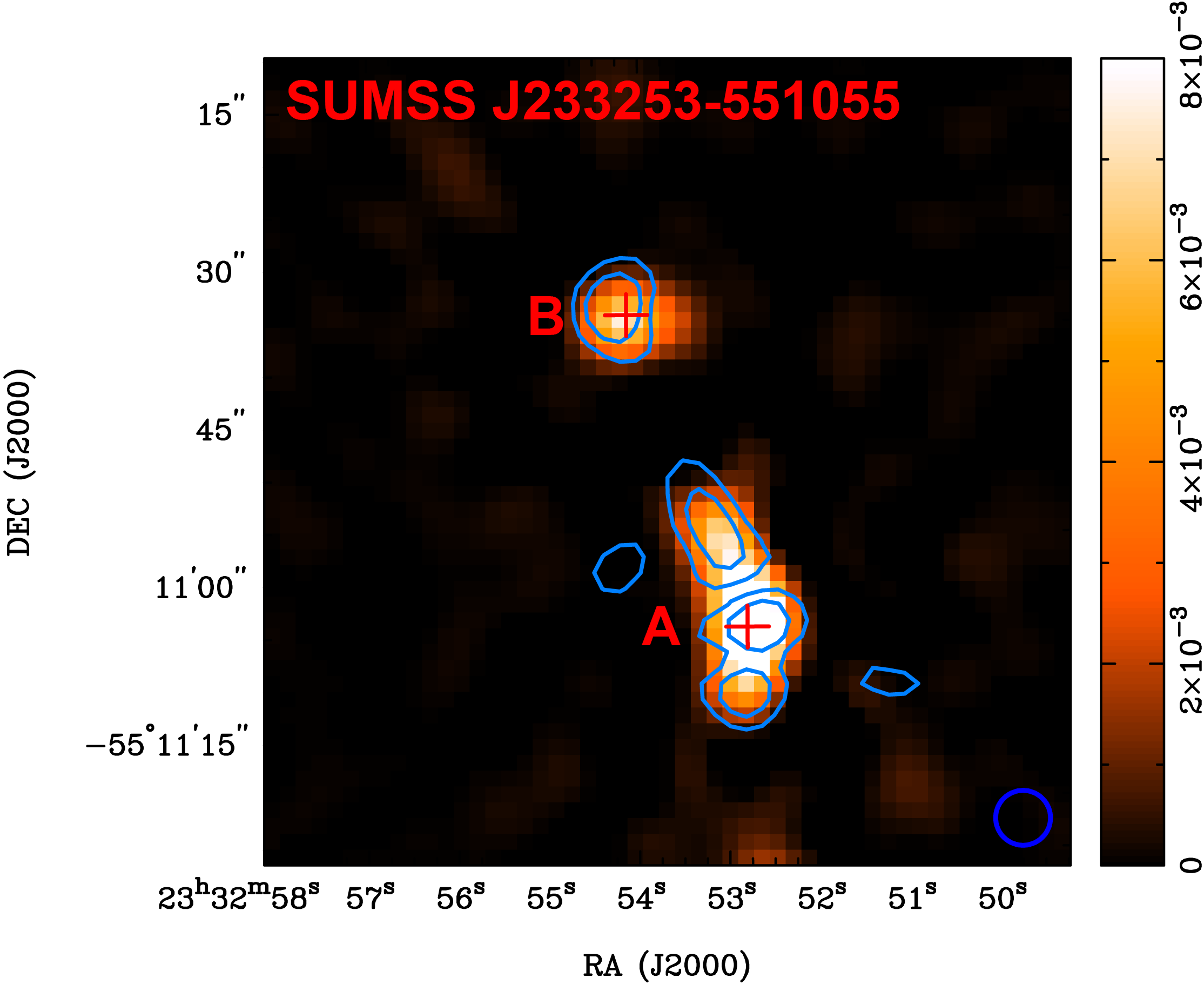}
\includegraphics[width=\columnwidth]{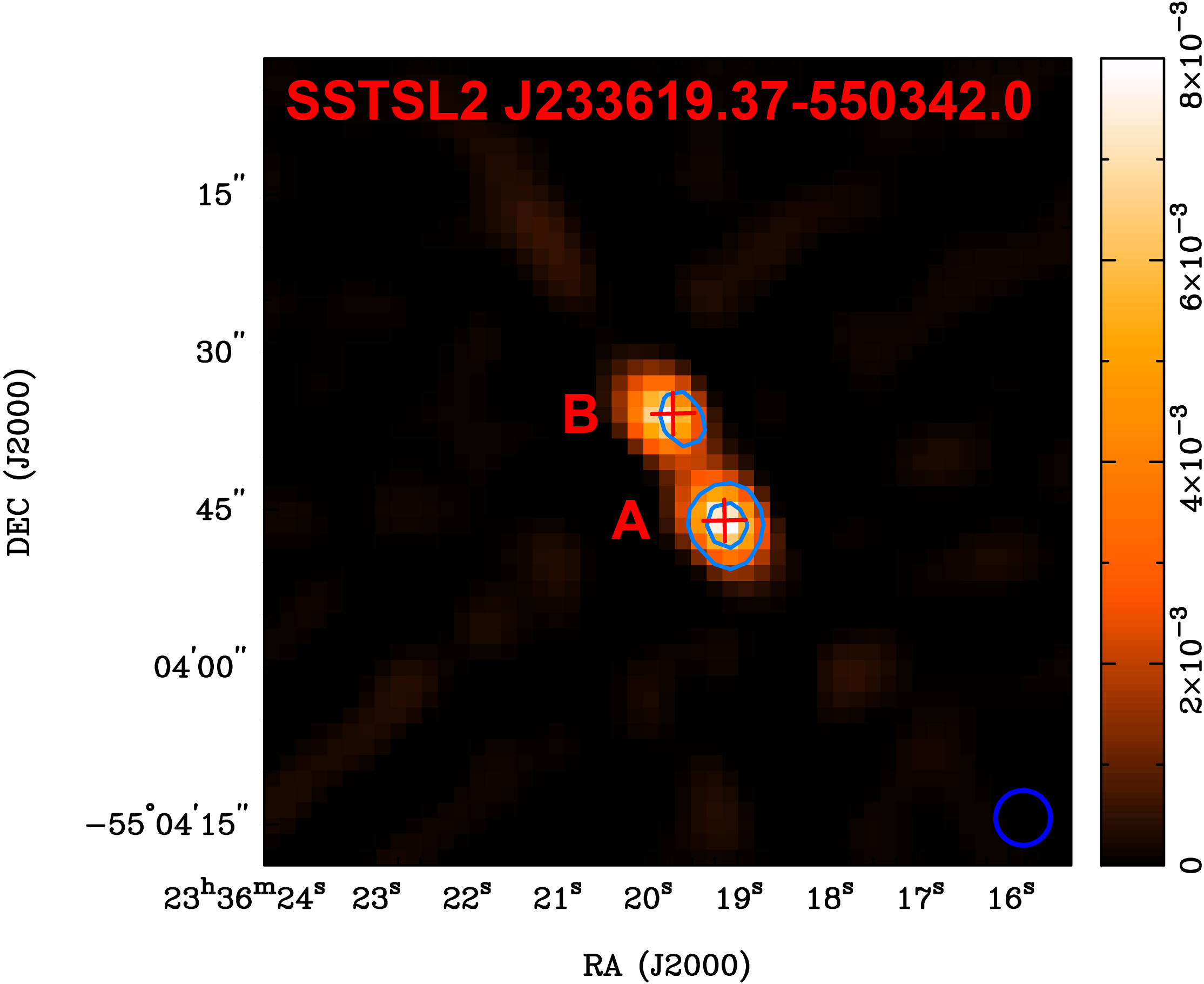}
\includegraphics[width=\columnwidth]{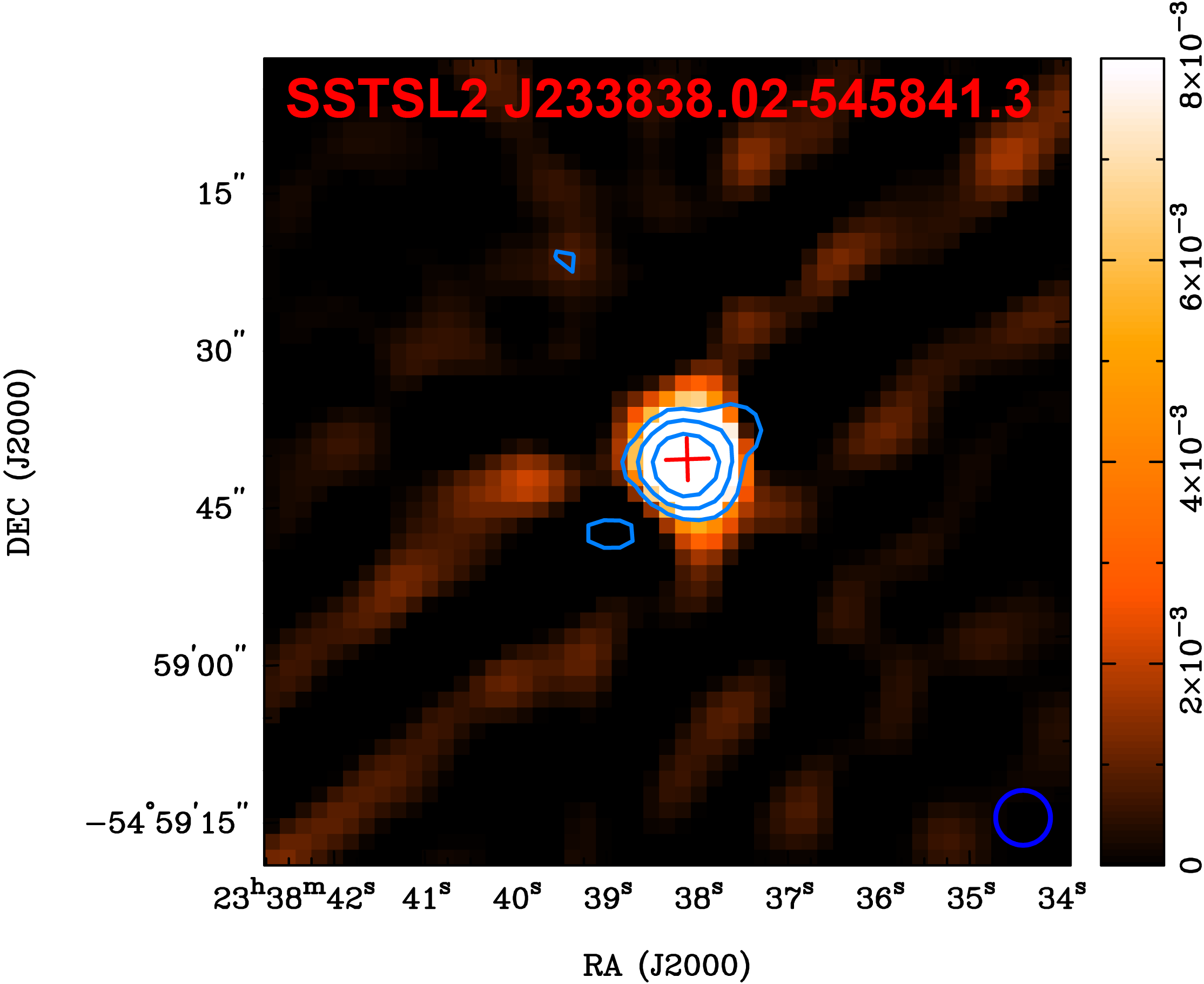}
\caption{continued.}
\end{figure*}

\end{document}